\documentclass[pre,twocolumn,endfloats,amsmath]{revtex4}

\usepackage{graphicx}
\usepackage{psfrag}

\newcommand{\eq}[1]{Eq.~(\ref{#1})}
\newcommand{\eqs}[1]{Eqs.~(\ref{#1})}
\newcommand{\fig}[1]{Fig.~\ref{#1}}
\newcommand{\app}[1]{Appendix~\ref{#1}}
\newcommand{\sect}[1]{Section~\ref{#1}}

\begin{document}
\title{Diffusive Transport in Networks Built of Containers and Tubes}

\author{L. Lizana and Z. Konkoli\footnote{Address: Chalmers University of
Technology, SE-412 96 G\"{o}teborg, Sweden. E-mail:
\texttt{zorank@fy.chalmers.se}. Tel. +46 31 772 3186; Fax: +46 31 41
69 84. \\}}

\affiliation{Department of Applied Physics, Chalmers University of
Technology and G{\"o}teborg University}

\date{\today}

\begin{abstract}
We developed analytical and numerical methods to study a transport of
non-interacting particles in large networks consisting of $M$
$d$-dimensional containers $C_1,\ldots,C_M$ with radii $R_i$ linked
together by tubes of length $l_{ij}$ and radii $a_{ij}$ where
$i,j=1,2,\ldots,M$.  Tubes may join directly with each other forming
junctions. It is possible that some links are absent. Instead of
solving the diffusion equation for the full problem we formulated an
approach that is computationally more efficient. We derived a set of
rate equations that govern the time dependence of the number of
particles in each container $N_1(t),N_2(t),\ldots,N_M(t)$. In such a
way the complicated transport problem is reduced to a set of $M$ first
order integro-differential equations in time, which can be solved
efficiently by the algorithm presented here.  
The workings of the method have been demonstrated on a couple of
examples: networks involving three, four and seven containers, and one
network with a three-point junction.  Already simple networks with
relatively few containers exhibit interesting transport behavior. For
example, we showed that it is possible to adjust the geometry of the
networks so that the particle concentration varies in time in a
wave-like manner. Such behavior deviates from simple exponential
growth and decay occurring in the two container system. 
\end{abstract}

\maketitle

\section{Introduction} \label{Introduction}

The goal of this work is to find a method that describes diffusive
transport of particles on a network built up of spherical containers
connected by tubes. It is possible that not all containers are
connected to each other and tubes may join together forming junctions
(without a container being present). In such a way one can generate an
enormous number of network topologies. One example is show in
\fig{MainNetwork}. The networks consists of $M$ containers
(reservoirs) labeled $C_1,\ldots,C_M$ of radii $R_i$ connected by
tubes of length $l_{ij}$ and radii $a_{ij}$ where $i,j=1,2,\ldots,M$.

Our work is motivated by experiments discussed in
refs.~\cite{Orwar,KSMDKO,KDKKBKJLHVO} and ~\cite{LPHR,HR,HSR}. The
first set of references describes how to create and manipulate
microscale sized compartments (vesicles) connected by nanotubes. These
structures can be applied in a number of ways \cite{KDKKBKJLHVO}. For
example, the non-diffusive (forced) transport was studied in
\cite{Orwar}. In this work we study different kind of transport were
only passive diffusion is allowed. Also, by including the reactions in
the theoretical setup one could describe biochemical reactions in a
milieu close to their natural habitat. This idea was pursued
experimentally in ref. \cite{KSMDKO}.

The second set of references deals with networks of chemical reactors
of macroscopic sizes connected to each other by tubes. The exchange of
reactants is mediated through the tubes and controlled by pumps. It
was shown that it is possible to use this device to carry out pattern
recognition tasks.  In making the device smaller the external pumping
can be removed and the transport can be limited to pure
diffusion. This kind of setup is close to the situation studied here.

For large networks, links connecting opposite sides of the network may
be rather long. Accordingly, one can not expect an exponential decay
of the number of particles in the containers and this is the situation
we are mostly interested in.  Obviously, to describe such a situation
one can attempt to solve the diffusion equation numerically and obtain
the distribution function $\rho(\vec{r},t)$ that describes how
particles spread throughout the network. 

Obtaining the solution of the diffusion equation for a complicated
geometry for a large networks gets highly impractical.  In this work
we develop a method of calculation that is computationally efficient
and can be used to study transport in large networks.  Instead of
finding the full distribution function $\rho(\vec{r},t)$ we introduce
a set of slow variables that capture the most important aspect of
particle transport, the number of particles in each container
$N_1(t),N_2(t),\ldots,N_M(t)$, and derive equations that describe how
they change in time. This is the central result of the paper.


A couple of related problems have been addressed previously in
refs. \cite{Grigoriev, Berezhkovskii, Dagdug, Dagdug2, Bezrukov}.
Escape of a particle through a small hole in a cavity was studied in
\cite{Grigoriev}. The work of \cite{Berezhkovskii} deals with the
problem of the hole connected to a short tube. The tube length and
hole radius are roughly of the same size, mimicking a cell membrane
having a thickness greater than zero. A couple of equilibration cases
have also been studied \cite{Dagdug,Dagdug2,Bezrukov}. The papers just
indicated treat the intra container dynamics in much more detail than
we do. In here, for simplicity reasons, the particle concentration is
assumed flat in the container and the validity of this approximation
is checked numerically in section \ref{IdealMixing}.  In our notation,
the studies \cite{Grigoriev, Berezhkovskii, Dagdug, Dagdug2, Bezrukov}
can be classified as $M=1,2$ cases. Our main interest is in networks
with large $M$.

This paper is organized as follows. In section \ref{ProblemDef} the
problem is defined and the general results are stated. The derivation
of the rate \eq{GeneralEq} is explained in sections \ref{SecEmpt} and
\ref{SecRate} where the emptying of a single container into a tube,
and emptying of a container into another container through the tube is
studied. The single exponential asymptotics of the two container
system and related first order rate equations are found and discussed
in section \ref{SingExpSec}. Up to this point only a two-container
system is treated while section \ref{SecGeneralEqs} deals with an
arbitrary network topology. Section \ref{IdealMixing} elaborates on
the assumption of well stirred containers. A numerical comparison to
the diffusion equation is made. In section \ref{CaseStudies} numerical
case studies of various network structures are performed. In
particular a three way junction and an example of a larger network are
studied. The summary and outline of future work is given in section
\ref{Conclusions}. Technical details are found in the
appendices. Appendix \ref{Numerical} describes the numerical procedure
used for solving the rate equations. The rate equations for the cases
studied in \sect{CaseStudies} are explicitly derived in
\app{CaseRates}. It can be shown that the presence of tube junctions
can be eliminated altogether from the dynamical equations when the
time is large. This is demonstrated in \app{SectJunction}.

\section{Problem definition and Main Result}\label{ProblemDef}
Describing the particle transport in a network depicted in
\fig{MainNetwork} is far from trivial and in order to solve the
problem several assumptions are made. We assume that (i) particles
move solely by diffusion (the fluid in which the particles move stands
still) and (ii) particles do not disturb each other. With these
assumptions the complicated dynamical problem at hand is reduced to
solving the time dependent diffusion equation:
\begin{equation}\label{DiffEq}
\partial_t \rho(\vec{r},t) = \nabla \cdot \left[D(\vec{r})\nabla
\rho(\vec{r},t)\right].
\end{equation}
Here $\rho(\vec{r},t)$ is the concentration (particle density) and
$D(\vec{r})$ is the diffusion coefficient which may be position
dependent. The walls are particle impenetrable
\begin{equation}
\partial_n \rho(\vec{r},t) = 0
\end{equation}
where $\partial_n \equiv \hat{n}\cdot \nabla$, and $\hat{n}$ is the
unit vector perpendicular to the wall. The total number of particles
is a conserved quantity.

Equation (\ref{DiffEq}) could in principle be solved numerically using
a brute force approach (e.g. the Finite Element Method or the Finite
Difference method). However, in sections \ref{SecEmpt} and
\ref{SecRate} we will show that it is possible to describe particle
transport in terms of a finite number of variables, the number of
particles in each container $N_1,...,N_M$. Also, it might be easier to
understand particle transport in such a setup. The dynamics of
$N_i(t)$ $i=1,\ldots, M$ is governed by
\begin{equation}\label{GeneralEq}\begin{array}{lll}
 \dot{N}_i(t) & = & \sum_{j=1}^M {\cal C}_{ji}\frac{V_{d-1}(a_{ji})}{V_d(R_j)}
                    \int_0^t dt'{\cal N}_j(t')\sigma_{ji}(t-t')\\\\

              &   & -\sum_{j=1}^M {\cal C}_{ij}\frac{V_{d-1}(a_{ij})}{V_d(R_i)}
            \int_0^t dt'{\cal N}_i(t')\left[\Delta_{ij}(t-t')\right.\\\\
          &   & \left. +\kappa_{ij}(t-t')\right]
\end{array}\end{equation}
where
\begin{equation}
{\cal N}_i(t) \equiv\dot{N}_i(t)+N_{i0}\delta(t)
\end{equation}
and $\delta(t)$ is the Dirac delta-function, \mbox{$\int_0^\infty
dt\,\delta(t) = 1$}. Here and in the following the dot over symbol
denotes time derivative. The connectivity matrix \mbox{${\cal C}_{ij} \in
\{0,1\}$} describes how the nodes are linked (note that \mbox{${\cal
C}_{ii}=0$)}, $a_{ij}$ is the radius of the tube (link) from $i$ to
$j$ and $V_d(R_j)$ is the volume of a $d$ dimensional sphere
\mbox{$V_d(r) = [2\pi^{d/2} /d\Gamma(d/2)]r^d$} corresponding to
container $j$ having radius $R_j$. $N_{i0}$ denotes
$N_i(t=0)$. Equation (\ref{GeneralEq}) is derived under the assumption
of ideally mixed containers. The rate coefficients are given by
\begin{equation}\label{GeneralRateCoef}\begin{array}{lll}
\Delta_{ij}(t) &= & \displaystyle\sqrt{\frac{D_{ij}}{\pi t}}\\\\

\kappa_{ij}(t) &= &\displaystyle2\sqrt{\frac{D_{ij}}{\pi
t}}\sum_{k=1}^\infty\exp\left(-\frac{(k \ell_{ij})^2}{D_{ij}t}\right)\\\\

\sigma_{ij}(t) &= &\displaystyle2\sqrt{\frac{D_{ij}}{\pi t}}\sum_{k=0}^\infty
\exp\left(-\frac{((2k-1) \ell_{ij})^2}{4D_{ij}t}\right).
\end{array}\end{equation}
$\ell_{ij}$ is the link length and $D_{ij}$ is the corresponding
diffusion coefficient. The theory is developed for the general
case where the diffusion constant in each tube may be different.

Figure \ref{MainNetwork} also shows the existence of tube
junctions. They are treated by \eq{GeneralEq} by letting the container
radius coincide with that of the tube. Since the tubes are initially
empty, so are junctions, $N_{i0}=0$.
Eqs. (\ref{GeneralEq})-(\ref{GeneralRateCoef}) are the central results
of this paper and their derivation is a major topic of the subsequent
sections.

\section{Emptying of a reservoir through an infinitely long tube} 
\label{SecEmpt}

To derive \eq{GeneralEq} we start with the simplest possible case and
consider particle escape from a container through an infinitely long
tube [see \fig{InfTubeFig}, panel (a)]. The main reason for this is to
show how to couple the dynamics of the tube and the container.  Also,
such setup captures the short time description of the full network
problem when the particles escaping the containers do not yet 'feel'
that the system is closed [the short time dynamics is contained in
$\Delta(t)$, see \sect{SecRate}].

The particle concentration $\rho(\vec{r},t)$ is governed by the
diffusion equation supplemented with the boundary conditions that the
walls are impenetrable and that $\rho(\vec{r},t)$ has to vanish for
$x\rightarrow \infty$. The concentration in the tube and in the
container are interwoven in a highly non-trivial way through what is
occurring at the tube opening. Given that the current density out of
the container $j(0,y,z,t)$ is known [see \fig{InfTubeFig} (a)] one
could solve the diffusion problem and find the concentration profile
in the container and the number of particles. Furthermore, one could
find a relationship
\begin{equation}\label{Functional}
\rho(0^-,y,z,t) = {\cal F} [j(0,y,z,t)]
\end{equation}
where $j(0,y,z,t) = -D\,\lim_{x\rightarrow 0^-}\hat{x}\cdot \nabla
\rho(x,y,z,t)$. ${\cal F}$ is a functional that we know exists but is
unlikely to be found in a closed analytic form. 

To find $j(0,y,z,t)$ it is assumed that the concentration in the
vicinity of the tube opening can be approximated
with
\begin{equation}\label{DeCoupling}
\rho(x,y,z,t) = f(y,z,t) c(x,t) \,\,\,\,\ x\gtrsim 0
\end{equation}
where $c(x,t)$ is a one-dimensional concentration and $f(z,y,t)$ is a
function that projects the value of $c(x,t)$ onto a radial direction.
Equation~(\ref{DeCoupling}) is valid for large $x$ when
$f(y,z,t)$ is constant but not in general case. Arbitrary density
profile at the opening will in time smear out due to radial
diffusion. 

By assumption, the concentration profile in the tube is governed by
$c(x,t)$ and it is coupled to the concentration in the container as
follows. Both concentration and current have to be continuous as one
moves from the container into the tube leading to
\begin{equation}\label{BC}
\rho(0,y,z,t) = f(y,z,t) c(0,t).
\end{equation}
and
\begin{equation}\label{BCFlow}
j(0,y,z,t) = f(y,z,t) \lim_{x\rightarrow 0^+} \frac{\partial}{\partial
x} c(x,t)
\end{equation}
Please note that $\lim_{x\rightarrow 0^+} \frac{\partial}{\partial
x}c(x,t)$ in \eq{BCFlow} is a functional of $c(0,t)$. Taking into
account particle conservation at the tube opening leads to
\begin{equation}\label{PartConservation}
\int_{x=0} dS \;\rho(0,y,z,t) = c(0,t)
\end{equation}
which after using \eq{BC} results in the condition $\int dS\, f(y,z,t)
= 1$. The problem has four unknowns $\rho(0,y,z,t)$, $j(0,y,z,t)$,
$c(0,t)$ and $f(y,z,t)$ and four equations (\ref{Functional}),
(\ref{BC})-(\ref{PartConservation}) and is fully defined. However, it
is not tractable in this complicated form and we proceed to simplify
it.

Instead of $\rho(\vec{r},t)$ a more useful variable is the total
number of particles in the container
$N(t)=\int_{x<0}dV\rho(\vec{r},t)$ governed by
\begin{equation}\label{CouplingInf}
\dot{N}(t) = -J(t)
\end{equation}
where $J(t)$ is the flow of particles that leave the container through
the tube opening $J(t) = \int dS j(0,y,z,t)$.  There are two special
cases where the current $J(t)$ can be determined analytically in terms
of container variables. (i) If the exit is a fully absorbing disk with
radius $a$, the concentration is always zero at the the interface
$\rho(0,y,z,t) = 0$. In \cite{BePu} it is shown that the current
$J_\infty$ through such disk when placed at an infinite otherwise
reflecting wall is $J_\infty = 4D_c a\rho_\infty$ where $\rho_\infty$
is the particle concentration at infinity and $D_c$ denotes the
diffusion constant in the container. A reasonable assumption for the
container, at least when the tube radius is smaller than the radius of
the container, is that the concentration profile far away from the
exit is flat and can approximately be taken to be $N(t)/V_d(R)$. Using
this for $\rho_\infty$ yields $J_\infty(t) = 4D_caN(t)/V_d(R)$. (ii)
If the opening is completely closed, the current is zero and
$\rho(0,y,z,t) = N(t)/V_d(R)$ [in such a case $N(t)= N(0)$]. A linear
interpolation between (i) and (ii) yields
\begin{equation}\label{LinearInterpol}
\rho(0,y,z,t) = \frac{N(t)}{V_d(R)}\left[1-\frac{J(t)}{J_\infty(t)}\right].
\end{equation}
Please note that in Eq.~(\ref{LinearInterpol}) $\rho(0,y,z,t)$ is
assumed constant across the interface. This approximation is verified
numerically in \fig{InletProfile} that shows $f(y,z,t)\approx
V_{d-1}(a)^{-1}$. Also, when $a\ll R$ one has $J(t)/J_\infty(t)\ll 1$
and second term in (\ref{LinearInterpol}) can be neglected.  This
approximation is verified by numerical calculations in
\sect{IdealMixing}.

At this point, the full problem has been mapped on to a very simple
geometry depicted in \fig{InfTubeFig} (b): a one dimensional line
(tube) connected to a point (container). Tube dynamics is
characterized by a one dimensional particle density $c(x,t)$ along the
line and all container dynamics, how complicated it may be, has been
projected on to a single variable $N(t)$.

The only part of the problem that remains to be solved is the
diffusion through the tube.  This part of the problem can be
approximated by one dimensional diffusion since $f(y,z,t)$ is
constant. The constant concentration profile at the tube opening
remains such in the tube interior (provided the tube radius does not
change along $x$ direction). With assumptions at hand the coupling
\eq{BC} becomes
\begin{equation}\label{BC2}
c(0,t) = V_{d-1}(a)\frac{N(t)}{V_d(R)}.
\end{equation}
The concentration profile along the tube [initially empty $c(x,0)=0$]
is given by the diffusion equation
\begin{equation}
\frac{\partial c(x,t)}{\partial t} = D \frac{\partial^2
c(x,t)}{\partial x^2}\;\;\;\;\;\;x\in[0,\infty)
\end{equation}
supplemented with boundary conditions according to \eq{BC2} and
$c(\infty,t)=0$. The solution can be found by the Laplace transform
method \cite{Schaum} and is given by
\begin{equation}\label{SolInf}
c(x,s)=c(0,s)\,e^{-x\sqrt{s/D}}.
\end{equation}
where $c(x,s)=\int_0^\infty dt \,c(x,t) e^{-st}$. Integrating
\eq{BCFlow} over the tube interface area at $x=0$ gives
\begin{equation}\label{FicksLaw}
J(t) = -D\lim_{x\rightarrow 0}\frac{\partial}{\partial x}c(x,t).
\end{equation}
Combining \eqs{CouplingInf}, (\ref{BC2}) and (\ref{FicksLaw}) leads to
a rate equation in the Laplace transform space
\begin{equation}\label{CouplingInfInS}\begin{array}{lll}
sN(s)-N_0 &= &\displaystyle-\lim_{x\rightarrow 0}\sqrt{sD}N(s)
\frac{V_{d-1}(a)}{V_d(R)}\, e^{-x\sqrt{s/D}}\\\\ &=
&\displaystyle-\sqrt{sD}N(s)\frac{V_{d-1}(a)}{V_d(R)}
\end{array}\end{equation}
where ${\cal L}[\dot{N}(t)]=sN(s)-N_0$. It is tempting to rewrite this
equation in the time domain in the form of convolution
\begin{equation}\label{GenRateEq}
\dot{N}(t) = -\int_0^t dt' k(t')N(t-t')
\end{equation}
representing a general form of a rate law, where
$k(t)=\frac{V_{d-1}(a)}{V_d(R)} {\cal L}^{-1}[\sqrt{Ds}]$. However,
this is impossible and can be seen in several ways.

First, $\sqrt{s}$ has no well defined inverse Laplace transform and
$k(t)$ that would enter into the rate equation (\ref{GenRateEq}) is
ill-defined. Second, this problem could possibly be resolved by
inverting \eq{SolInf} to obtain $c(x,t)$ and inserting the result into
\eq{FicksLaw} which leads to
\begin{equation}\label{Illegal}
J(t) \propto -\lim_{x\rightarrow 0}\frac{\partial}{\partial x}\int_0^t
            dt' \,\frac{x}{t'^{3/2}}\, e^{-x^2/4Dt'} N(t-t').
\end{equation}
In general $N(t)$ is unknown. To evaluate the expression above in a
way that would result in a rate equation involves interchanging
derivation and integration.  This is only allowed if the integral is
uniformly convergent in the interval $x\in[0,\infty)$
\cite{LePage}. It is easy to see from \eq{Illegal} that this is not
the case and the interchange is illegal.  Another possibility is to
use partial integration but this strategy does not work since one ends
up with non-convergent integrals as $x\rightarrow 0$. Thus,
\eq{GenRateEq} does not exist for a semi-infinite case. Also, it is
intuitively clear that one can not observe pure exponential decay
since the system is infinite.

For an infinite system an asymptotic rate law of the type $\dot{N}(t)
\propto -N(t)$ simply does not exist. This can also be seen from the
exact expression for $N(t)$ which can be obtained from
\eq{CouplingInfInS} by solving for $N(s)$ and finding the inverse
Laplace transform \cite{Schaum}
\begin{equation}\label{SolInfInT}
N(t) = N_0\exp\left[ Dt \left(\frac{V_{d-1}(a)}{V_d(R)}
\right)^2\right] {\rm erfc} \left[\sqrt{Dt} \frac{V_{d-1}(a)}{V_d(R)}
\right].
\end{equation}
$N_0$ is the initial number of particles in the container. Using
approximation erfc$(z)\approx \frac{e^{-z^2}}{\sqrt{\pi}} \frac{1}{z}$
for large $z$ gives $N(t) \propto \frac{1}{\sqrt{t}}$.

Due to the complications discussed above the rate equation has to be
stated in terms of $\dot{N}(t)$
\begin{equation}\begin{array}{lll} \label{AlmostRate}
\dot{N}(t) &= &-\displaystyle \frac{V_{d-1}(a)}{V_d(R)}\, {\cal L
}^{-1}\left[sN(s) \sqrt{\frac{D}{s}}\right]\\\\
&= &-\displaystyle\frac{V_{d-1}(a)}{V_d(R)} \int_0^t dt' {\cal
N}(t') \Delta(t-t')
\end{array}\end{equation}
where
\begin{equation}
 {\cal N}(t)\equiv {\cal L }^{-1}\left[sN(s)\right] = \dot{N}(t) +
N_0\delta(t)
\end{equation}
and
\begin{equation}\label{DefOfDelta}
\Delta(t) \equiv {\cal L}^{-1}\left[\sqrt{\frac{D}{s}}\right] =
\sqrt{\frac{D}{\pi t}}.
\end{equation}
Please note that it is impossible to rewrite the right hand side of
\eq{AlmostRate} in such a way that it would solely involve dependence
on $N(t)$. When the system is closed (e.g. by adding another
container) the situation changes.

\section{A Rate Equation for a Two Container System} \label{SecRate}
The system under consideration consists of a one dimensional rod
parameterized by $a$ and $\ell$ connected to two ideally mixed
containers having radii $R_1$ and $R_2$ depicted in
\fig{TwoContSytem}.

The diffusion equation for the tube [initially empty, $c(x,t) = 0$]
connected to the two containers at $x=0$ and $x=\ell$ is given by
\begin{equation}
\frac{\partial c(x,t)}{\partial t} = D\frac{\partial^2
c(x,t)}{\partial x^2}\;\;\;\;\;\;x\in(0,\ell)
\end{equation}
with boundary conditions analogous to \eq{BC2}
\begin{equation}
\begin{array}{lr}
c(0,t) = N_1(t)\frac{V_{d-1}(a)}{V_d(R_1)}, & c(\ell,t) =
N_2(t)\frac{V_{d-1}(a)}{V_d(R_2)}.
\end{array}\end{equation}
The solution in Laplace transform space is given by
\begin{equation}\label{SolInS}
\displaystyle c(x,s) = \frac{V_{d-1}(a)}{V_d(R_1)} sN_1(s) \phi_1(x,s) +\frac{V_{d-1}(a)}{V_d(R_2)}
sN_2(s)\phi_2(x,s)
\end{equation}
where
\begin{equation}\begin{array}{lr}
\phi_1(x,s) =  \frac{\sinh\left((x-\ell)\sqrt{s/D}\right)}
{s\sinh\left(\ell\sqrt{s/D}\right)}, &

\phi_2(x,s) =  \frac{\sinh\left(x\sqrt{s/D}\right)}
{s\sinh\left(\ell\sqrt{s/D}\right)}.
\end{array}\end{equation}
Matching the current at both ends as in \eq{FicksLaw}
\begin{equation}\begin{array}{llr}
\dot{N}_1(t) &=& D\,\lim_{x\rightarrow 0}\frac{\partial}{\partial
x}c(x,t)
\\\\
\dot{N}_2(t) &=& - D\,\lim_{x\rightarrow \ell}\frac{\partial}{\partial
x}c(x,t)

\end{array}\end{equation}
yields a set of rate equations in Laplace transform space for the two
container system
{\small
  \begin{subequations}
  \label{RateInS}
  \begin{eqnarray}
  sN_1(s)-N_{10} & = & -sN_1(s)\frac{V_{d-1}(a)}{V_d(R_1)}
                       \sqrt{\frac{D}{s}}\frac{\cosh\left(\ell\sqrt{s/D}\right)}
                       {\sinh\left(\ell\sqrt{s/D}\right)}\nonumber\\
                 & &   +sN_2(s)\frac{V_{d-1}(a)}{V_d(R_2)}\sqrt{\frac{D}{s}}
                       \frac{1}{\sinh\left(\ell\sqrt{s/D}\right)}\nonumber\\
                 & &\\
  sN_2(s)-N_{20} & = & -sN_2(s)\frac{V_{d-1}(a)}{V_d(R_2)}
                       \sqrt{\frac{D}{s}}\frac{\cosh\left(\ell\sqrt{s/D}\right)}
                       {\sinh\left(\ell\sqrt{s/D}\right)}\nonumber\\
                 &   & +sN_1(s)\frac{V_{d-1}(a)}{V_d(R_1)}\sqrt{\frac{D}{s}}
                       \frac{1}{\sinh\left(\ell\sqrt{s/D}\right)}.\nonumber\\
                 & &
  \end{eqnarray}
  \end{subequations}
}
The terms having minus signs are the outflow while the ones having
plus signs represent the inflow. A closer look at the the rate
equations above reveals an important link to the semi-infinite case:
the outflow is proportional to $\sqrt{s}$ for large $t$ (small
$s$). In other words, the semi-infinite case is recovered as a short
time expansion of \eqs{RateInS} (see \sect{SecEmpt}). The physical
interpretation is that initially, the particles feel as if they were
entering an infinitely long tube. This implies that the outflow term
can be divided into two parts reflecting this observation
\begin{equation}\begin{array}{l}\label{DeltaSigma}
  \sqrt{\frac{D}{s}}\left[\frac{\cosh\left(\ell\sqrt{s/D}\right)}
  {\sinh\left(\ell\sqrt{s/D}\right)}\right] \equiv \Delta(s) + \kappa(s).
\end{array}\end{equation}
$\Delta(s)$ is taken from \eq{DefOfDelta} and controls the short time
dynamics that resembles the one of the semi-infinite
system. $\kappa(s)$ describes the asymptotic long time behavior when
the particles are 'aware' of the existence of another side and can be
found from \eq{DefOfDelta} and \eq{DeltaSigma}
\begin{equation} \label{RateCoefInS1}
   \kappa(s) = \sqrt{\frac{D}{s}}\frac{\exp\left(-\ell\sqrt{s/D}\right)}
               {\sinh\left(\ell\sqrt{s/D}\right)}.
\end{equation}
The inflow rate is labeled $\sigma(s)$
\begin{equation}\label{RateCoefInS2}
    \sigma(s) = \sqrt{\frac{D}{s}}\frac{1}{\sinh\left(\ell\sqrt{s/D}\right)}.
\end{equation}
The inverse Laplace transforms of $\Delta(s)$, $\kappa(s)$ and
$\sigma(s)$ are shown in \eq{GeneralRateCoef} and depicted in
\fig{RateGraph}. The long time behavior of $\kappa(t)$ and $\sigma(t)$
can be found from a small $s$ expansion of the \eqs{RateCoefInS1} and
(\ref{RateCoefInS2})
\begin{equation} \begin{array}{lll}\label{RateLimit2}
  \kappa(t) &\approx & \frac{D}{\ell} - \sqrt{\frac{D}{\pi t}}\\
  \sigma(t) &\approx & \frac{D}{\ell}
                        \left[
                              1 -\exp\left(-\frac{6Dt}{\ell^2}\right)
                        \right].
\end{array}\end{equation}
Taking the limit $t \rightarrow \infty$ yields
\begin{equation}\begin{array}{lr}\label{RateLimit}
\;\;\sigma(\infty) = \frac{D}{\ell}\;\;
&\;\;\kappa(\infty) = \frac{D}{\ell}.
\end{array}\end{equation}

It is more convenient to express the rate \eqs{RateInS} in time domain
{\small
  \begin{subequations}
  \label{RateEqTwoCont}
  \begin{eqnarray}
  \dot{N}_1(t) & = & \frac{V_{d-1}(a)}{V_d(R_2)} \int_0^t dt'
                     {\cal N}_2(t')\sigma(t-t')\nonumber\\
               &   & -\frac{V_{d-1}(a)}{V_d(R_1)}\int_0^t dt'
                     {\cal N}_1(t')\left[\Delta(t-t')
                     + \kappa(t-t')\right]\nonumber\\
               & &\\
  \dot{N}_2(t) & = & \frac{V_{d-1}(a)}{V_d(R_1)} \int_0^t
                     dt'{\cal N}_1(t')\sigma(t-t')\nonumber\\
               &   & -\frac{V_{d-1}(a)}{V_d(R_2)}\int_0^t dt'
                     {\cal N}_2(t')\left[\Delta(t-t')+\kappa(t-t')\right]
                      \nonumber\\
               & &
  \end{eqnarray}
  \end{subequations}
}
$\kappa(t)$ and $\sigma(t)$ are not present in the rate equation for
the semi-infinite case (see Eq. \ref{AlmostRate}) and arise only when
the system is finite.

A numerical solution to \eqs{RateEqTwoCont} is shown in
\fig{NumSolTwoCont} (see \app{Numerical} for a more elaborate
discussion regarding the numerical procedure). The number of particles
decays exponentially which is verified in \fig{LogTwoCont} where the
straight line attained after some time is the evidence of a single
exponential decay. The figure also shows a non-exponential regime for
small $t$, described by $\Delta(t)$. Terms proportional to $\Delta(t)$
are in the following referred to as $\Delta$-terms.

For small $t$ particles rush into the tube with a large current. At
$t=0$ the current is infinite, $\lim_{t\rightarrow 0}
\dot{N_1}(t)=\infty$. Thus, exactly at $t=0$ it is impossible to
define the exit rate from container and such situation extends to any
other time instant. In \emph{strict} mathematical sense, it is
impossible to define exit rate from the container for any $t>0$ (when
the concentration profile at the tube opening is different from
zero). This can be illustrated on a simple example.  Let $V$ be a
volume divided into two sub-volumes $V_1$ and $V_2$ such that $V_1$
and $V_2$ touch each other and exchange particles by diffusion.  The
dynamical variables of interest are the total number of particles in
each sub-volume $N_1(t)$ and $N_2(t)$. The goal is to derive some kind
of rate equation for $N_1(t)$ and $N_2(t)$. We focus on the flow from
$V_1$ to $V_2$. In a small time interval $\epsilon$ one will have
$N_1(t+\epsilon)\propto N_1(t)(1-\alpha\sqrt{\epsilon})$ and
$N_2(t+\epsilon)\propto N_2(t)+\beta N_1(t)\sqrt{\epsilon}$, where
$\alpha$ and $\beta$ are numerical constants. The effective exchange
rate that describes the flow from $V_1$ into $V_2$ is given by
\begin{equation}
   k_{21}(t)=\lim_{\epsilon\rightarrow 0}
             \frac{N_2(t+\epsilon)-N_2(t)}{N_1(t)\epsilon}
             \propto\lim_{\epsilon\rightarrow 0} \epsilon^{-1/2}
\end{equation}
which is infinite.  At any time instant $t$ an infinite amount of
particles (per unit time) is rushing from $V_1$ into $V_2$ (and vise
versa). However, the infinite flows from $V_1$ to $V_2$ and the other
way around cancel each other out resulting in a finite net flow giving
smooth curves for $N_1(t)$ and $N_2(t)$. $t=0$ is special since there
is no counter flow from $V_2$ to $V_1$ which explains why
$\lim_{t\rightarrow 0} \dot{N}_1(t) = \infty$.

The non-exponential regime grows with tube length $\ell$ and might
play a significant role in studying transport processes in networks
having long connections.  For large times $\kappa(t)$ and $\sigma(t)$
start to dominate and one observes exponential decay. It will be shown
in \sect{SingExpSec} how to derive rate equations that describe this
regime.

\section{Analysis of the general rate equation: Emergence of
         a single exponential solution} \label{SingExpSec}
It is intuitively clear that in the case of the two-node network
discussed in previous section one should have an exponential decay
(growth) for the number of particles in the container $C_1$ ($C_2$)
(see Figs. \ref{NumSolTwoCont} and \ref{LogTwoCont}): asymptotically
the time dependence of $N_1(t)$ and $N_2(t)$ is given by
\begin{equation}
   N_{1,2}(t) = N_{1,2}(\infty) + {\cal A}_{1,2} \exp(-t/\tau)
\end{equation}
where $\tau^{-1}$ is the decay exponent that governs the late time
asymptotics and $ {\cal A}_{1,2}$ is the amplitude of decay.  This
fact is not easily predicted from the form of the general rate
equation given in (\ref{GeneralEq}). To understand the emergence of
such behavior a more thorough investigation of \eq{TwoContRate} is
needed.

To obtain the exact value of the decay exponent one has to study the
structure of poles of $N_{1,2}(s)$.  The poles fully determine the
form of $N_{1,2}(t)=\sum_{p=0}^\infty a_p e^{s_pt}$ where $a_p$ is the
residue of $N_{1,2}(s)$ at pole $s_p$. The values of $N_{1,2}(\infty)$
are determined by the $p=0$ term ($s_0=0$). The exponential decay rate
is determined by
\begin{equation}\label{tau_s1}
  \tau^{-1} = -s_1
\end{equation}
Rewriting \eqs{RateInS} in matrix form yields
\begin{equation}\label{DefofM}
s\vec{N}(s) - \vec{N}_0 = {\cal M}\vec{N}(s)
\end{equation}
where
\begin{equation}\begin{array}{lr}
\vec{N}(s) = [N_1(s),N_2(s)]^T, & \vec{N}_0 = [N_{10},N_{20}]^T
\end{array}\end{equation}
and
\begin{equation}\begin{array}{l}\label{DetM}
{\cal M} =\frac{qD}{\ell^2}
   \left[
         \begin{array}{cc}
                   -\frac{V_{\rm  tube}}{V_d(R_1)} \coth q
                 &  \frac{V_{\rm  tube}}{V_d(R_2)}\frac{1}{\sinh q}\\\\

                    \frac{V_{\rm  tube}}{V_d(R_1)}\frac{1}{\sinh q}
                 & -\frac{V_{\rm  tube}}{V_d(R_2)}\coth q
         \end{array}
   \right]
\end{array}\end{equation}
with $q^2 = s\ell^2/D$ and $V_{\rm tube} = V_{d-1}(a)\ell$. The poles
are calculated from $\det (s-{\cal M})=0$ since \mbox{$\vec{N}(s) =
(s-{\cal M})^{-1}\vec{N_0}$} and $(s-{\cal M})^{-1} \propto 1/\det
(s-{\cal M})$. Evaluating $\det (s-{\cal M}) = 0$ gives
\begin{equation}\label{TransEq}
q^2 + q\left[\frac{V_{\rm tube}}{V_d(R_1)} +
   \frac{V_{\rm tube}}{V_d(R_2)}\right]\coth q +
   \frac{V_{\rm tube}^2}{V_d(R_1)V_d(R_2)} = 0
\end{equation}

Equation (\ref{TransEq}) is a transcendental equation and has many
solutions $q_p$ that determine the value of the poles $s_p = q_p^2
D/\ell^2$ where $p=1,2,\ldots,\infty$ and in particular
\begin{equation}\label{q_s1}
  s_1= \frac{q_1^2D}{\ell^2}
\end{equation}
that, together with \eqs{tau_s1}, gives a relationship between $q_1$
and the decay rate $\tau^{-1}=-q_1^2D/\ell^2$. Note that $q^2$ has
been factored out from \eq{TransEq} and $s_0=0$ ($q_0=0$) is the
additional pole that determines the values of $N_{1,2}(\infty)$. Also,
note that $q_1$ depends only on parameters describing geometry of the
network.

Apart from determining the structure of the poles,
\eqs{DefofM}-(\ref{DetM}) are a good starting point for classifying
various schemes for obtaining approximative forms of the rate
\eqs{RateEqTwoCont}. Equations (\ref{RateEqTwoCont}) do not have the
form of the general rate law stated in \eq{GenRateEq} (due to the
presence of the $\Delta$-terms). Such a rate law would be easier to
understand intuitively. For example, the emergence of the single
exponential decay could be seen more easily in \eq{GenRateEq} than in
\eqs{RateEqTwoCont}. Also, an approximative form might be easier to
implement numerically, though at the cost of a lower accuracy at the
end.  The idea is to perform a small $s$ expansion of \eq{DefofM}
based on a desired accuracy. In here we consider two cases.\\

\noindent{\bf (a) Lowest order expansion:} Performing the expansion
\begin{equation}\begin{array}{ll}
  q \coth q \approx 1 & \ \ \ \ \ \frac{q}{\sinh q} \approx 1
\end{array}\end{equation}
of ${\cal M}$ in \eq{DetM} leads to a matrix that is constant, and taking the
inverse Laplace transform of \eq{DefofM} gives the following set of
rate equations
\begin{equation}\label{DagdugRateEqn}
  \dot{N}_1(t) = -\dot{N}_2(t) =
                 V_{\rm tube}\frac{D}{\ell^2}\left[\frac{N_2(t)}{V_d(R_2)} -
                \frac{N_1(t)}{V_d(R_1)} \right].
\end{equation}
These equations were already stated in ref.~\cite{Dagdug}. It is
interesting to see that they emerge as a special case of the scheme
discussed here. Also, \eq{DagdugRateEqn} can be obtained by following
another route. Performing partial integration of \eqs{RateEqTwoCont}
with terms containing $\Delta(t)$ omitted leads to the exactly the
same form of rate equations as given in (\ref{DagdugRateEqn}). This
procedure is discussed below.

The $\Delta$-terms are only present when the system is infinite. Since
the problem is finite, terms proportional to $\Delta(t)$ will be
sub-leading for large $t$. This can be seen from a small $s$ (large
$t$) expansion of \eqs{DefOfDelta} and ({\ref{RateCoefInS1})
\footnote{A small $s$ expansion of \eq{DefOfDelta} and
(\ref{RateCoefInS1}) in Laplace transform space yields ${\cal
L}^{-1}[\Delta(s)] \propto t^{-1/2}$ and ${\cal L}^{-1}[\kappa(s)]
\propto {\rm const}$.  Since $\Delta(t)$ and $\kappa(t)$ always
combine in a sum, $\kappa(t)$ will dominate the outflow for large
$t$.}.  Also, partial integration of the $\Delta$-terms is impossible
since the derivative of $\Delta(t)$ is proportional to $t^{-3/2}$ and
diverges when $t \rightarrow 0$.

Omitting $\Delta$-terms in \eqs{RateEqTwoCont}, and performing partial
integration leads to
{\small
  \begin{subequations}
  \label{TwoContRate}
  \begin{eqnarray}
     \dot{N}_1(t) &=& \frac{V_{d-1}(a)}{V_d(R_2)}
                      \int_0^t dt' N_2(t-t')
                      \dot{\sigma}(t')\nonumber\\
                  & & -\frac{V_{d-1}(a)}{V_d(R_1)}
                      \int_0^t dt'
                      N_1(t-t')\dot{\kappa}(t')\\
    \dot{N}_2(t) &=& \frac{V_{d-1}(a)}{V_d(R_1)}\int_0^t dt'
                     N_1(t-t')\dot{\sigma}(t')\nonumber\\
                 & & -\frac{V_{d-1}(a)}{V_d(R_2)} \int_0^t dt'
                     N_2(t-t')\dot{\kappa}(t')
  \end{eqnarray}
  \end{subequations}
}
Since $\dot{\sigma}(t)$ is peaked for small $t$ (see \fig{RateGraph}),
the contribution to the integrals (convolutions) stems mainly from
small values of $t'$ which justifies the following approximation
\begin{equation}\label{intapprox}
 \int_0^t dt' N_{1,2}(t-t')\dot{\sigma}(t') \approx N_{1,2}(t) \int_0^t dt'
 \dot{\sigma}(t')= N_{1,2}(t)\sigma(t).
\end{equation}
where $\sigma(0)=0$ was used. The same applies for $\kappa(t)$. Using
\eq{intapprox} in (\ref{TwoContRate}) leads to
{\small
  \begin{subequations}
  \label{FirstAttempt}
  \begin{eqnarray}
    \dot{N}_1(t) & = & \frac{V_{d-1}(a)}{V_d(R_2)}
                     N_2(t)\sigma(t) - \frac{V_{d-1}(a)}{V_d(R_1)}
                     N_1(t)\kappa(t) \ \ \ \ \ \ \ \\
    \dot{N}_2(t) & = & \frac{V_{d-1}(a)}{V_d(R_1)}
                     N_1(t)\sigma(t) - \frac{V_{d-1}(a)}{V_d(R_2)}
                     N_2(t)\kappa(t). \ \ \ \ \ \ \
  \end{eqnarray}
  \end{subequations}
}
Inserting $\kappa(\infty)$ and $\sigma(\infty)$ found in
\eq{RateLimit} into (\ref{FirstAttempt}) yields
(\ref{DagdugRateEqn}). This example shows how the $\Delta$-terms
disappears from the description when the system is finite. However,
contrary to the partial integration method, the expansion of
\eq{DefofM} gives a more systematic and controlled approach.

Equation (\ref{DagdugRateEqn}) is simple and computationally
efficient. It could be easily used to describe large
networks. However, it has several drawbacks that can be
identified. The solution to \eq{DagdugRateEqn} is given by
\begin{equation} \label{DagdugApprox}
  N_{1,2}(t)-N_{1,2}(\infty) =
                      \left[N_{1,2}(0)-N_{1,2}(\infty)\right]
                      \exp\left(-t/\tau_{1,a}\right).
\end{equation}
The decay rate $\tau_a^{-1}= -q_{1,a}^2D/\ell^2$ is determined by
\begin{equation}\label{Rate1}
  q_{1,a}^2 = -\frac{V_{\rm tube}[V_d(R_1) +  V_d(R_2)]}{V_d(R_1)V_d(R_2)}.
\end{equation}
The number of particles in each container as $t\rightarrow \infty$ is
\begin{equation}\label{Ninfty}
 \frac{N_1(\infty)}{V_d(R_1)} = \frac{N_2(\infty)}{V_d(R_2)}=
 \frac{N_{10} + N_{20}}{V_d(R_1)+V_d(R_2)}.
\end{equation}
Equation (\ref{Ninfty}) is not correct. The correct values for
$N_{1,2}(\infty)$ are given by
\begin{equation} \label{LimitDist}
  \frac{N_1(\infty)}{V_d(R_1)} = \frac{N_2(\infty)}{V_d(R_2)}=
  \frac{N_{10} + N_{20}}{V_d(R_1)+V_d(R_2)+V_{\rm tube}}
\end{equation}
The discrepancies between \eq{Ninfty} and (\ref{LimitDist}) become
increasingly important for long tubes which are likely to occur in
large networks. For example, in a case where the tube and reservoir
volumes are equal, \eq{Ninfty} predicts $N_1(\infty) =
N_2(\infty)=N_{tot}/2$, $N_{tot}=N_{10} + N_{20}$, while the exact
result from \eq{LimitDist} is $N_{tot}/3$. The particle decay exponent
given in \eq{Rate1} only holds when $V_{\rm tube} \rightarrow 0$. It
strongly deviates from the exact value when the tube is long (see
\fig{TransEqFig}). \\

\noindent{\bf (b) Higher order expansion:} Using the expansion
\begin{equation}\begin{array}{ll}\label{Ordo}
 q \coth q \approx  1+\frac{q^2}{3} &\ \ \ \ \  \frac{q}{\sinh q} \approx  1
\end{array}\end{equation}
for ${\cal M}$, inserting in \eq{DefofM} and taking inverse Laplace
transform, leads to a set of rate equations (given in
\app{appFirstTauM2}) that are unsatisfactory due to the following
reasons. First, they predict a spurious jump in $N_{1,2}(t)$ as $t
\rightarrow 0$, and the limiting values $N_{1,2}(\infty)$ are not
correct. Second, the rate exponent that results from these equations
is not that accurate. This particular example shows that it is
important to have a balanced expansion for elements of ${\cal M}$. For
example, instead of expanding $q\coth q$ directly one has to expand
$\sinh q$ and $\cosh q$ separately in such a way that same powers in
the nominator and denominator are obtained. When this strategy is
followed a much better approximation is obtained as shown bellow.

The next order expansion, gives correct limits for $N_{1,2}(t)$ when
$t\rightarrow 0$ and $t\rightarrow\infty$ and leads to a relatively
accurate value for the decay exponent. Using the expansion
\begin{equation}\begin{array}{ll}
  q\coth q \approx  \frac{1+q^2/2}{1+q^2/6}\ \ \ \ \
  \frac{q}{\sinh q} \approx  \frac{1}{1 +q^2/6}
\end{array}\end{equation}
in ${\cal M}$ gives the following set of equations
{\small
  \begin{subequations}
  \label{ThirdAttempt}
  \begin{eqnarray}
  \dot{N}_1(t)  &=&  -N_1(t)\frac{3D}{\ell^2}\frac{V_{\rm tube}}{V_d(R_1)}\nonumber \\
                & &  +\frac{12D^2}{\ell^4}\frac{V_{\rm tube}}{V_d(R_1)}
                     \int_0^t dt' N_1(t-t')\exp\left(-\frac{6Dt'}{\ell^2}\right) \nonumber\\
                & &  -\frac{6D^2}{\ell^4}\frac{V_{\rm tube}}{V_d(R_2)}\nonumber
                     \int_0^t dt' N_2(t-t')\exp\left(-\frac{6Dt'}{\ell^2}\right)\nonumber\\
                & &\\
  \dot{N}_2(t)  &=&  -N_2(t)\frac{3D}{\ell^2}\frac{V_{\rm tube}}{V_d(R_2)}\nonumber\\
                & &  \frac{12D^2}{\ell^4}\frac{V_{\rm tube}}{V_d(R_2)}\nonumber
                     \int_0^t dt' N_2(t-t')\exp\left(-\frac{6Dt'}{\ell^2}\right)\nonumber\\
                & &  -\frac{6D^2}{\ell^4}\frac{V_{\rm tube}}{V_d(R_1)}
                     \int_0^t dt' N_1(t-t')\exp\left(-\frac{6Dt'}{\ell^2}\right)\nonumber\\
                & &
\end{eqnarray}
\end{subequations}
}
Solving $\det (s-{\cal M}) = 0$ to get hold of the decay exponent in
analytical form becomes in this case rather tedious since finding the
value $q_1$ amounts to finding a root of a fourth degree
polynomial. The calculation simplifies somewhat if equal container
volumes are considered, $V_d(R_1)=V_d(R_2)\equiv V_d$. In such a case
one has
\begin{equation}\begin{array}{lll}\label{Rate3}
 q_{1,b}^2 &=&-\frac{1}{2V_d}\left[3(2V_d+V_{\rm tube})\right.\\\\

           & &-\left.\sqrt{3(12V_d^2-4V_dV_{\rm tube} + 3V^2_{\rm tube})}\right]
\end{array}\end{equation}

The main findings of this section are summarized in
Figs. \ref{NumSolTwoCont}, \ref{LogTwoCont} and
\ref{TransEqFig}. Figures \ref{NumSolTwoCont} and \ref{LogTwoCont}
depict a numerical solution of \eq{RateEqTwoCont} (solid line)
compared with the approximations discussed in this section for a case
where the tube and container volumes are equal. Figure
\ref{TransEqFig} shows a detailed analysis of the decay rate.

For very short times there is a difference between \eqs{ThirdAttempt}
and (\ref{RateEqTwoCont}) in \fig{NumSolTwoCont} and
\ref{LogTwoCont}. These arise due to the partial elimination of
$\Delta$-terms.  For example, \eq{ThirdAttempt} does not predict
$\lim_{t\rightarrow 0} \dot{N}_1(t)=\infty$ [\fig{NumSolTwoCont}, the
dashed line lies above the solid line near $t=0$ for curves depicting
$N_1(t)$]. This is the reason why the curve for $N_1(t)$ obtained from
\eq{ThirdAttempt} underestimates the emptying of container $C_1$. This
effect is more pronounced for the $N_1(t)$ coming from
\eq{DagdugApprox}. There, the $\Delta$-terms are eliminated altogether
[\fig{NumSolTwoCont}, the dotted line depicting $N_1(t)$ lies above
solid and dashed lines].  Also, Figure \ref{NumSolTwoCont} shows that
there is a large error in $N_{1,2}(\infty)$ for curves obtained by
\eqs{DagdugRateEqn}.

In \fig{LogTwoCont} the natural logarithm of $[N_1(t) -
N_1(\infty)]/N_{tot}$ is shown. For short times the dynamics is not
exponential but after some time a straight line is attained which is
the evidence of single exponential behavior. The curve corresponding
to \eq{DagdugApprox} (dotted line) does not predict the correct decay
exponent which is manifested in a different slope. The decay exponent
predicted by \eq{ThirdAttempt} is a better estimate for the decay
rate: the slopes of the dashed and solid lines more or less
coincide. This fact is shown more clearly in \fig{TransEqFig}.

Figure \ref{TransEqFig} depicts the dependence of $q_1^2$ as a
function of the tube volume.  The numerical solution to \eq{TransEq}
(solid line), which gives the exact value of the decay exponent, is
compared to the values of $q^2_{1,a}$ (dotted line) and $q^2_{1,b}$
(dashed line). All three cases work well when $V_{tube}\rightarrow
0$. As the tube volume increases $q_{1,a}^2$ deviates more and more
from $q_1^2$. The same holds for $q_{1,b}^2$, though its value lies
much closer to $q_1^2$. For example, when all volumes are equal $q_1
=-1.71$ and $q^2_{1,b} = -1.63$ while $q^2_{1,a} = -2$.

In this section methods of finding rate equations and decay exponents
for the two container problem was deduced.  Figures
\ref{NumSolTwoCont}, \ref{LogTwoCont} and \ref{TransEqFig} show that
for increasing tube volumes the rate equations given in
\eq{DagdugApprox} are not capable of describing the dynamics.  The
approximation given in \eq{ThirdAttempt} works better.  For short time
dynamics none of the developed methods are valid and the full rate
\eq{GeneralEq} is the only alternative.  In the subsequent section all
rate equations discussed up to this point will be extended to work for
any network structure.

\section{The General Expression}\label{SecGeneralEqs}
In this section results and methods obtained and developed for
two-node network will be extended to work for any network
structure. The complete dynamics for the two-node network is
formulated in \eq{RateEqTwoCont}. For an arbitrary network the outflow
(OUT) from container $i$ to container $j$ is proportional to
$\Delta_{ij}(t)+\kappa_{ij}(t)$ and the inflow (IN) from container $j$
to container $i$ is proportional to $\sigma_{ji}(t)$:
{\small
  \begin{subequations}
  \label{GenEq}
  \begin{eqnarray}
  {\rm OUT}_{i\rightarrow j} & = & \frac{V_{d-1}(a_{ij})}{V_d(R_i)} \int_0^t dt'
                                   {\cal N}_i(t')\left[\Delta_{ij}(t-t')\right.
                       \nonumber\\
                              &&   \left.+ \kappa_{ij}(t-t')\right]\\
  {\rm IN}_{j\rightarrow i}  &= &  \frac{V_{d-1}(a_{ji})}{V_d(R_j)} \int_0^t dt'
                                   {\cal N}_j(t')\sigma_{ji}(t-t').
  \end{eqnarray}
  \end{subequations}
}
This implies
\begin{equation} \label{GeneralSum}
  \dot{N}_i(t) = \sum_{j\ne i} {\cal C}_{ij}\left[{\rm IN}_{j\rightarrow i}
               - {\rm OUT}_{i \rightarrow j}\right],\;\;\;\; i = 1,\ldots, M
\end{equation}
where ${\cal C}_{ij}$ is the conductivity matrix discussed in
\sect{ProblemDef}. The final result is stated in \eq{GeneralEq}.

It was argued earlier that if the tube volume is small a very simple
first order rate equation can be stated. It is found in
\eq{DagdugRateEqn}. Extending this equation to the case of arbitrary
topology yields
\begin{equation}\label{LargeVApprox}
\dot{N}_i(t) = \sum_{j\ne i} {\cal C}_{ij}V_{d-1}(a_{ij})\frac{D_{ij}}{\ell_{ij}}
\left[\frac{N_j(t)}{V_d(R_j)}-\frac{N_i(t)}{V_d(R_i)}\right].
\end{equation}
Note the symmetry relations $a_{ij}= a_{ji}$, $\ell_{ij}= \ell_{ji}$
and $D_{ij}= D_{ji}$.

If a more sensitive solution is desired, a set of equations of the
type (\ref{ThirdAttempt}) is suggested. An extension of this equation
is shown below
{\small
  \begin{subequations}
  \label{SensitiveRate}
  \begin{eqnarray}
  {\rm OUT}_{i\rightarrow j} &= &  N_i(t)\frac{3D_{ij}}{\ell_{ij}}
                                   \frac{V_{d-1}(a_{ij})} {V_d(R_i)}\nonumber\\
                             & &-  \frac{12D_{ij}^2} {\ell_{ij}^3}\nonumber
                                   \frac{V_{d-1}(a_{ij})}{V_d(R_i)} \int_0^t dt'
                                   \exp\left(\frac{-6Dt'}{\ell_{ij}^2}\right)
                   N_i(t-t')\\
                 & &\\
  {\rm IN}_{j\rightarrow i}  & = & \frac{6D_{ji}^2} {\ell_{ji}^3}
                                   \frac{V_{d-1}(a_{ji})}{V_d(R_j)}\int_0^t dt'
                                   \exp\left(\frac{-6D_{ji}t'}{\ell_{ji}^2}
                   \right)N_j(t-t').\nonumber\\
  \end{eqnarray}\end{subequations}
}
Inserting \eq{SensitiveRate} in \eq{GeneralSum} yields an
approximative form of rate equations for an arbitrary network.

In previous section we investigated differences between \eqs{GenEq},
(\ref{LargeVApprox}) and \eq{SensitiveRate} using the two-node network
as a study case ($M = 2$). It is expected that the findings of
previous section also apply for larger networks. This analysis is not
conducted here. Having general expressions at hand more complicated
network structures can be investigated. However, the assumption of
well stirred containers will be discussed first since it is crucial
for the derivation of the rate equations~(\ref{GeneralEq}).

\section{The assumption of ideally mixed containers}\label{IdealMixing}
An ideally mixed container has no concentration gradients. If a
particle has entered, it can be found anywhere within the compartment
with equal probability. In reality, a diffusing particle examines the
compartment in a random walk fashion until an opening is found. There
it has a possibility to escape and change the concentration. If the
tube radius is small ($a/R\ll 1$), a significantly longer time is
required to find an opening and escape than to examine the majority of
the compartment. The time needed to examine the majority of the
compartment is called mixing time and is given by $\tau_{{\rm mix}} =
\frac{R^2}{D}$. The time of finding a specific place or target having
radius $a$ is given by $\tau_{{\rm target}} = \frac{R^2}{D}
\frac{R}{a}$ \cite{Stange}. For cubic compartments the radius of the
sphere $R$ is replaced by the edge length, in this case $2R$ (see
\fig{SolDiffEq}). Thus, we argue that if $a\ll R$ then $\tau_{{\rm
mix}} \ll \tau_{{\rm escape}}$ and the containers can be considered
ideally mixed at all times
\footnote{$\tau_{\rm target}$ is derived under the assumption of a
fully absorbing target of radius $a$. This is not a correct
description of a tube opening since it allows reentry. This estimate
serves however as a worst case scenario.}. This is supported by
numerics.

Figure \ref{SolDiffEq}~(b) shows a numerical solution of the diffusion
equation in two dimensions in a geometry depicted in
\fig{SolDiffEq}~(a). The solution clearly shows that the assumption of
well stirred containers becomes very good for $a/R \ll 1$, even for
skewed initial distributions. The solution to the diffusion equation
was found using a standard implicit finite difference discretization
method \cite{NumRes}.

In \fig{SolDiffEq}~(b) one observes a systematic discrepancy: the
assumption of ideally mixed containers tends to overestimate the decay
rate. This derives from the fact that the assumption of well stirred
containers over estimates the number of particles at the tube inlet
leading to a larger exit rate.

\section{Case studies}\label{CaseStudies}
Up to this point the two node network was the only example discussed.
It was used as elementary building block when constructing the rate
equation (\ref{GeneralEq}). Such network is rather simple and does not
offer any spectacular behavior. In this section more complicated
structures will be studied that are shown in Figs. \ref{panel} and
\ref{LargeNetwork}: Two realizations of a three-node network are
studied first, case 1 and 2, depicted in \fig{panel}~(a) and (b). They
possess one more level of complexity than the two-node network shown
in \fig{TwoContSytem}. Case 3, \fig{panel}~(c), is a four-node network
that has a T-shape structure. Case 4, \fig{panel}~(d), has the shape
of a star and involves a tube junction. It will be shown that the
transport properties of this structure can be controlled in such way
that it might serve as a diffusion-based transistor. These three and
four node-networks, despite being rather simple, exhibit a large
variety of outputs. Case 5, \fig{LargeNetwork}, demonstrates the
transport properties for larger networks that are impossible to
predict without using \eq{GeneralEq}. When solving \eq{GeneralEq}
numerically the containers and the tubes are considered three
dimensional. Moreover, the diffusion coefficients $D_{ij}$ and the
tube radii $a_{ij}$ were kept the same for all tubes. All graphs are
scaled with the total number of particles in the system $ N_{tot} =
\sum_{i = 1}^M N_i(0)$.\\

%
\noindent {\bf Case 1: The Line.} Three reservoirs are lined up on a
straight line as depicted in \fig{panel}~(a). The transport equations
for this system can easily be written down using \eq{GeneralEq} and
are listed in \app{LineRate}. A numerical solution is shown in
\fig{LineGraph}. This three-node system exhibits a new characteristic
that can not be found in the two node network. The curve depicting the
number of particles in the middle container (solid line) has a
maximum (an extremum) point.

The extremum point is a manifestation of an unbalance between inflow
and outflow in the middle container. This unbalance derives from an
asymmetry in the structure and arise only when $\ell_{12} < \ell_{23}$
or $V_1<V_3$. This provides a possibility of designing the output
pattern through simple geometrical changes in the structure. Fig.
\ref{LineTiming} is a simple demonstration of this design possibility
where $\ell_{12}$ is varied so that the arrival time of the maximum of
$N_2(t)$, depicting the number of particles in the middle compartment
$C_2$, is changed. The picture also shows that an increase in
$\ell_{12}$ results in both an increased arrival time and a wider
peak. The height of the maximum is controlled by the value $V_2$. A
decrease in $V_2$ suppresses the peak and vice versa. \\

%
\noindent {\bf Case 2: The Triangle.}  The connectivity of the line is
changed by adding an extra link between containers $C_1$ and $C_3$ so
that it forms the shape of a triangle, as shown in
\fig{panel}~(b). The rate equations are derived in \app{TriangleRate}
and a numerical solution is shown in \fig{Triangle2}. The initial
distribution of particles, see the inset in \fig{Triangle2}, is chosen
in such a way that a minimum is produced in the curve depicting the
number of particles in container $C_2$.

The extremum point can be enhanced or removed completely in the same
way as was demonstrated for case 1. Such minimum will be absent unless
the geometry and initial distribution of particles are tailored in a
specific way. In general, such sensitivity of geometrical changes and
changes in the location where particles are injected is observed for
all cases studied in this section.

The triangle structure studied here exhibits shorter equilibrium
process than the linear structure considered previously [see
\fig{panel}~(a)]. In the case of a triangle structure particles can
spread more efficiently due to the additional routing possibility
$C_1\rightarrow C_3$.\\

%
\noindent {\bf Case 3: The T-Network.} It is interesting to see how
the behavior of the cases 1 and 2 changes when a new node is added to
the network. In here we study a situation where an extra node $C_4$ is
connected to container $C_2$ in the structure depicted in
\fig{panel}~(a). In such a way one gets a T-shaped (star) network
shown in \fig{panel}~(c). This alternation of structure leads to a
significant change in behavior, as shown in \fig{LineGraphTop}. When
compared to the cases 1 (one maximum) and 2 (one minimum) the curve
depicting $N_2(t)$ exhibits an additional extremum point: both minimum
and maximum are present simultaneously.  The right inset in
\fig{LineGraphTop} emphasizes this fact.

This scheme could be carried out further adding on more and more
reservoirs and adjusting the lengths and the initial distribution so
that the peaks arrive in consecutive order, possibly produce an
wave-like behavior. However, since the spread of the peaks increases
with increasing tube length, it might be numerically quite difficult
to see when the the extremum points occur or even if they actually
exist.\\

%
\noindent {\bf Case 4: The Junction.}  The next interesting network to
consider is the one with a junction present as depicted in
\fig{panel}~(d). A particular example of a three way junction is
studied. The network is built up by three reservoirs and three
tubes. The ends of the tubes coincide to form a junction. To obtain
the transport properties of such a network we start from the structure
studied in case 3 shown in \fig{panel}~(c). The junction is obtained
by reducing the radius of container $C_2$ in the middle until its
radius is roughly equal to the radii of the surrounding tubes, see
\fig{Junction1}. The rate equations describing the junction properties
have the same form as the equations describing case 3 (listed in the
appendix \ref{Trate}) with the substitution $V_d(R_3)\rightarrow
V_d(a)$. The equations are not given in order to save space.

A numerical solution of the rate equations for the junction is shown
in \fig{Junction}. This structure allows control of the particle flow
between $V_1$ and $V_3$ by adjusting the volume $V_2$ and the length
$\ell_{24}$ (see \fig{Junction1}). This setup could function as a
diffusion based transistor. For example, by making $\ell_{24}$ shorter
than $\ell_{34}$, compartment $C_2$ will initially attract diffusing
particles from $C_1$ to a greater extent than $C_3$ causing a time
delay in the particle arrival into container~$C_3$.

This fact is illustrated in \fig{Junction} where the maximum in the
curve for the number of particles in container $C_2$ (solid line)
indicates an initial accumulation of particles in $C_2$:
$N_2(t)/N_{tot}$ rises from 0 to 0.5 in the interval $t=0$ to
$Dt/\ell^2=5$. In this interval container $C_2$ accumulates the
majority of the particles released from container $C_1$. After the
peak has been reached the particles accumulated in $C_2$ are released
into $C_3$: $N_2(t)/N_{tot}$ continuously drops from value of 0.5
after $Dt/\ell^2=5$. By changing $\ell_{24}$ the curve for $N_2(t)$
can be manipulated exactly in the same way as done in
\fig{LineTiming}, but such analysis is not repeated.

For large networks involving junctions it might be desirable to
decrease the number of equations required to solve the transport
problem. It is demonstrated in \app{SectJunction} that the presence of
the junctions can be eliminated altogether when investigating dynamics
in the $t \rightarrow \infty$ regime for structures having large
container volumes ($a\ll R_i$, $i=1,\ldots,M$). Equations
(\ref{RateEqsJunction21})-(\ref{RateEqsJunction23}) show this
explicitly for the three node junction studied here.

The four cases studied up to now show that the transport dynamics is
very sensitive to geometrical changes and to the locations where
particles are injected. Small variations in tube lengths and container
volumes lead to unpredictable changes in curves depicting the time
dependence of the number of particles in each container
(Figs. \ref{LineGraph}, \ref{Triangle2}, \ref{LineGraphTop} and
\ref{Junction}).  Also, the shapes of the curves differs significantly
from a single exponential decay .\\

%
\noindent {\bf Case 5: Large Network.} Following the methods described
in this paper one could easily study diffusive transport in networks
containing hundreds or thousands of containers, tubes and
junctions. The computational cost scales linearly with both the number
of containers and number of tubes (assuming full connectivity of the
containers). We do not show explicit example with such large number of
containers.  To demonstrate the power of the method we study a more
pedagogical example, the case of a network that is built up by seven
containers, 9 tubes and a 4-way junction. In contrast to the previous
cases shown in \fig{panel} it is impossible to predict the transport
behavior of such a network without a numerical calculation.

Figure \ref{NetworkPanel}, panels~(a)-(c), shows numerical solutions
of \eq{GeneralEq} for the network depicted in
\fig{LargeNetwork}. Different initial distributions are used and are
introduced into inset of all figures. The darker the container appears
the more particles are injected into it. The dynamics is evidently
quite complex and all possible characteristics that were forced upon
the other cases are present. There is an exponential growth and decay
as well as curves having one or more extremum points. The different
transport behavior shown in \fig{NetworkPanel} stem only from
different initial conditions. If the structure no longer remained
fixed even more complicated patterns could be produced, only the
imagination sets the limits.

\section{Concluding remarks} \label{Conclusions}

We introduced a generic model for a diffusive particle transport in
large networks made of containers and tubes. The diffusion equation
that describes the distribution of particles $\rho(\vec{r},t)$
throughout the network was taken as a starting point. Instead of
calculating $\rho(\vec{r},t)$ explicitly, we followed another route
and developed a theoretical technique to solve the transport problem
using finite number of variables $N_1(t),\ldots,N_M(t)$ that describe
the number of particles in each container. First, a set of rate
equations was derived for the two-node network and second, they were
generalized to work for an arbitrary network structure. In such a way
we obtained the rate equations that govern dynamics of
$N_1(t),\ldots,N_M(t)$. These equations are summarized in
\eq{GeneralEq} and are the central result of the paper.

The transport equations were found by study the exchange of chemicals
between the container and the tube. It was demonstrated in
\sect{SecEmpt} how to couple the dynamics in the containers and tubes,
as stated in Eqs. (\ref{Functional}) and
(\ref{BC})-(\ref{PartConservation}). However, the coupling is too
complicated to be carried out in practice in the original
form. Several approximations were made in order to make such scheme
doable.

The tubes were assumed to be one dimensional lines (see Eq. \ref{BC}),
and the transport in the tubes was described in terms of a one
dimensional diffusion problem involving the concentration profile
along the line $c(x,t)$. The expression for $c(x,t)$ can be found
analytically using e.g. Laplace transform technique. In such a way the
tubes were eliminated from the problem.

We have considered diffusive non-interacting point particles which is
a plausible assumption for dilute solutions. One could consider the
situation when the particles disturb each other. In that respect, the
region of the tube interior is the most critical since the tubes can
be very narrow and exclusion effects will be mostly pronounced in
there. Such effects could be added into the theoretical description by
using results obtained from studies on diffusion with exclusion in one
dimensional systems~\cite{BD,JSI,EFCm}. We expect a very different
transport behavior when the diameters of the tubes becomes comparable
in size to that of the particles.

The dynamics in the container is not tractable analytically but it was
argued that when the tube radii are smaller than any other length
scales in the system (e.g. tube lengths or container radii) the
containers can be treated as ideally mixed at all times. This
assumption was verified numerically in \sect{IdealMixing}, and
simplifies all intra-container dynamics to one dynamical variable: the
total number of particles in a given compartment.

Evidently much of the container dynamics has been neglected but the
coupling is formulated in such a way that a more detailed description
can be developed should there be a need for that. For example, the
container dynamics could be better treated by using the techniques
presented in e.g. \cite{Dagdug}, or by further exploration of the
coupling equations (\ref{Functional}) and
(\ref{BC})-(\ref{PartConservation}). For example, one could keep the
second term in the right hand side of \eq{LinearInterpol} and use
$\rho(0,y,z,t) = N(t)/V_d(R)-J(t)/4D_ca$ instead of $\rho(0,y,z,t) =
N(t)/V_d(R)$. This procedure would lead to similar rate equations as
presented here with different forms for $\Delta(t)$, $\kappa(t)$ and
$\sigma(t)$.

Initially the particles feel as if they are escaping from the
container into an infinitely long tube. In this regime the number of
particles in the container decays non-exponentially. We have
identified terms in the rate equations describing this behavior: all
terms in \eq{GeneralEq} proportional to $\Delta(t)$ dominate when time
is small. This non-exponential regime grows with increasing tube
length and crossover time for this regime scales as $\ell^2/D$. Also,
terms containing $\Delta(t)$ contain solely dependence on
$\dot{N}_i(t)$ and can not be rewritten in the form that would involve
$N_i(t)$. Accordingly, it it impossible to rewrite \eq{GeneralEq} so
that it adopts a form of a general rate law (see
Eq. \ref{GenRateEq}). These issues were discussed in \sect{SecEmpt}.
The bottle neck lies in the definition of transport rate which, in
principle, is an ill-defined quantity (see discussions at the end of
sections \ref{SecEmpt} and \ref{SecRate}).

Eventually, for large times the decay is exponential. In such regime
term proportional to $\Delta(t)$ can be neglected in the rate
\eq{GeneralEq}. Other terms proportional to $\kappa(t)$ and
$\sigma(t)$ can be rewritten in the form of a general rate law,
leading to \eqs{LargeVApprox} and (\ref{SensitiveRate}). We showed
that \eq{LargeVApprox} is a special case of the general approximation
scheme developed in \sect{SingExpSec}: starting from rate
\eq{GeneralEq} in Laplace transform space we developed a series of
approximations that can be used to systematically describe the
asymptotic regime, resulting in \eq{SensitiveRate}. Also, we have
developed a procedure that can be used to eliminate junction points
for large times which reduces the number of variables further (see
\app{SectJunction}).

Already simple case studies that were used to illustrate the workings
of the method exhibit interesting behavior. For example, one can
identify three types of curves that appear in the plots depicting the
time dependence of the particle number in each container
(Figs. \ref{LineGraph} - \ref{LineGraphTop} and \ref{Junction}).
Type~I curves occur for the two node network. These lack extremum
points and the particle number either strictly rises or drops to
saturate to asymptotic values. Type II curves have one maximum or
minimum, and type III curves can have more and these are the most
interesting. Type II and type III curves normally describe the
particle number for the container in the network interior.

The existence of type I curves for large networks suggests that it
could be possible to understand the transport between two nodes in
terms of an effective two node network where a complicated structure
of links and containers in between two nodes is mapped onto an
effective link connecting them.  One can raise a more general
question: what is the smallest network that would have the same
transport properties as some sub-structure of a given large network?
For example, if one is interested in only three nodes of the structure
depicted in \fig{LargeNetwork}, e.g. $C_5$, $C_7$ and $C_8$, is there
a star-type network, e.g. as the one depicted in \fig{panel} (c) or
(d), that would have equivalent transport properties?

The presence of type II and type III curves indicates the possibility
that there might be curves that posses a larger number of extremum
points.  These are likely to occur in larger networks.  We can of
course manually enhance certain properties as height and width of
peaks. This will however become more and more complicated as the
network size increases. Problem is that the peaks that occur later
have larger width and might be harder to see.  For design purposes, it
is therefore necessary to build a learning mechanism or a search
engine, on top of our existing software, to select certain
characteristics in the curves depicting time evolution of
$N_1,...,N_M$.  Furthermore, to exploit such effects one has to
amplify them in some way. At the moment our study deals with transport
only, but reactions in the containers can be included as well, and
they could be tailored to amplify such effects (e.g. by choosing a
reaction of enzymatic type).

The techniques developed in this work can be used to study transport in
various systems. In the following we give a few examples.

(i) Although we exclusively study transport, our model can serve as a
platform for reaction-diffusion-based bio-computing devices
~\cite{SSFA, KK, Rambidi, CZ1, CZ2, CZ3, Ji, SM,LPHR, HR,
HSR,Akingbehin, AkinCon, Conrad, KirCon1, KirCon2, KamCon1, KamCon2}.
In particular, our work is applicable to studies of the
reaction-diffusion neuron ~\cite{Akingbehin, AkinCon, Conrad, KirCon1,
KirCon2, KamCon1, KamCon2}.  The reaction-diffusion neuron is an 2D
array of compartments that exchange chemicals by diffusion.

(ii) A large number of processes happening in the 
cell are governed by transport of
reactants and chemical reactions.  In order to avoid a need for
excessive storage facilities the chemical compounds are routed in an
orderly fashion between various places within and between the cells
and the chemical components arrive exactly at the right place at the
right time \cite{Stange,HM}. The setup in \fig{MainNetwork} captures
this aspect of the cell interior.

(iii) The transport on abstract mathematical networks (nodes and
links) has been studied extensively~\cite{Watts,WattsBook}. Such
studies are geometry free with emphasis on the topology of the network
graph (the connectivity patter, the average number of neighbors
etc.). The techniques developed in this work could be used to account
for the fact that the links between the nodes have physical length and
the transport along the links is not instantaneous.

In summary, the work we have presented is a step towards understanding
the transport properties of large networks where geometrical concepts
such as the length of the tubes play an important role.  The setup
employed in this study is rather simple. In order to be able to focus
on issues related to transport, the reactions are totally omitted. The
concentration profile in containers is assumed flat and this is good
approximation when tubes are thin.  Already simple examples of
networks we studied show a number of interesting properties. For
example, transport properties of the networks exhibit large
sensitivity to the geometrical changes in the structure. Also, one can
adjust structure to obtain wave-like behavior (with one or two
extremum points) in the curves that depict number of particles in
containers. When the complexity of the network increases one can
expect even more complicated behavior with larger number of extrema
points. The setup we use is generic and it is possible to expand the
model in many ways, e.g. by improving description of intra-container
dynamics, incorporating reactions, or allowing particles to disturb
each other.  It will be an interesting problem to try to explore these
questions further.

\begin{acknowledgments}

We would like to thank Prof. Owe Orwar and Prof. Mats Jonson for
fruitful discussions. The financial support of Prof. Owe Orwar is
greatly acknowledged.

\end{acknowledgments}


\appendix

\section{Numerical considerations}\label{Numerical}
In this section the numerical solution of \eq{GeneralEq} is
discussed. In \sect{SecEmpt} and \sect{SecRate} it was shown that the
time derivatives $\dot{N}_i(t),\ldots,\dot{N}_M(t)$ can be infinite at
$t=0$ which may cause numerical difficulties. However, the singular
part of the derivative can be factored out by making the substitution
\begin{equation}\label{Substitution}
\dot{N}_i(t)=-\left(\pi t\right)^{-1/2}\psi_i(t),\;\;\;\;\;i=1,\ldots,M
\end{equation}
where $\psi_i(t)$ is a smooth function of time. For small $t$ the
particles in the containers do not (yet) 'feel' the presence of
another side and behave as if entering an infinitely long tube. An
expression describing the transport behavior for for such a case was
found analytically and is stated in \eq{SolInfInT}. The time
derivative of \eq{SolInfInT} is proportional to $t^{-1/2}$ for small
$t$. Inserting \eq{Substitution} in \eq{GeneralEq} yields an equation
for $\psi_i(t)$ that was used for numerical calculations
\begin{equation}\label{GeneralEq2}\begin{array}{lll}
\psi_i(t)&=& \sum_{j=1}^M {\cal C}_{ji}\frac{V_{d-1}(a_{ji})}{V_d(R_j)}
             \left[\sqrt{\frac{t}{\pi}} \int_0^t \frac{dt'}{\sqrt{t'}}
             \psi_j(t')\sigma_{ji}(t-t')\right.\\\\
         & & \left.+\sqrt{\pi t} \,N_{j}(0)\sigma_{ji}(t)\right]\\\\

         & & -\sum_{j=1}^M {\cal C}_{ij}\frac{V_{d-1}(a_{ij})}{V_d(R_i)}
             \left[\sqrt{\frac{t}{\pi}}\int_0^t\frac{dt'}{\sqrt{t'}}
             \psi_i(t')\kappa_{ij}(t-t') \right. \\\\
         & & +\frac{\sqrt{D_{ij}t}}{\pi}\int_0^t\frac{dt'}
             {\sqrt{(t-t')}}\psi_i(t')\\\\
         & & \left. +\sqrt{\pi t}\,N_i(0) \left(\Delta_{ij}(t) +
             \kappa_{ij}(t)\right)\right].
\end{array}\end{equation}
Note that the expression for $\Delta_{ij}(t)$ has been inserted into
the integrals. Combining \eq{SolInfInT} and (\ref{Substitution}) leads
to
\begin{equation}
\psi_i(t) = N_{i0}-(\pi t)^{1/2}N_i(t),\;\;\;\;\;i=1,\ldots,M.
\end{equation}
and sets the initial condition $\psi_i(0)=N_{i0}$.

From \eq{GeneralEq2} two types of integrals can be identified:
\begin{equation}\begin{array}{lll}
 I_1^{t_n}\left[\psi\right] &=
&\int_0^{t_n}\frac{dt}{\sqrt{t'(t_n-t')}}\, \psi(t') \\\\
I_2^{t_n}\left[\psi\right] &=
&\int_0^{t_n}\frac{dt}{\sqrt{t'}}\, \psi(t')
\end{array}\end{equation}
where $\psi(t)$ is non-singular in the range of integration. The
quadrature formulas derived to solve $I_1^{t_n}\left[\psi\right]$ and
$I_2^{t_n}\left[\psi\right]$ are based on the methods described in
\cite{Mohamed}. The idea is that the singular part of the integrand,
$t'^{-1/2}$ and $[t'(t_n-t')]^{-1/2}$ respectively, is treated
exactly while the smooth part $\psi_i(t')$ is linearly interpolated
between $t_i$ and $t_{i+1}$. The resulting quadrature formulas are of
the form
\begin{equation}
I^{t_n}_i\left[\psi\right] \approx \sum_{j=0}^n w_{nj} \,\psi(t_j)
\;\; i=1,2
\end{equation}
where $w_{nj}$ are weights, $n = 1,2\ldots$ and $t_n = nh$. This
quadrature formula becomes exact when $\psi(t)$ is piecewise
linear. Calculating weights for $I_1^n[\psi]$ yields
{\small
\begin{equation}\begin{array}{lll}
w_{n0}&= &\sqrt{n-1} - (n-2) \arcsin \sqrt{\frac{1}{n}}\\\\
w_{ij}&= &\sqrt{(j-1)(n-j+1)}+\sqrt{(j+1)(n-j-1)}\\\\
      & &-2\sqrt{j(n-j)} + 2\left(j+1-\frac{n}{2}\right)\left[\arcsin\sqrt{\frac{j-1}{n}}\right.\\\\
      & &-2\left.\arcsin\sqrt{\frac{j}{n}} +\arcsin\sqrt{\frac{j+1}{n}}\right]\\\\
w_{nn}&=&\sqrt{n-1}+\pi\left(1-\frac{n}{2}\right)+(n-2)\arctan\sqrt{n-1}.
\end{array}\end{equation}
}
Calculating weights for $I_2^n[\psi]$ yields
\begin{equation}\begin{array}{ll}
w_{n0} &=\frac{4h^{1/2}}{3} \\\\
w_{nj} &=\frac{4h^{1/2}}{3}\left[(j+1)^{3/2}-2j^{3/2}+(j-1)^{3/2}\right]\\\\
w_{nn} &=\frac{2h^{1/2}}{3}\left[n^{3/2}-3(n-1)^{1/2}+2(n-1)^{3/2}\right].
\end{array}\end{equation}

Finally, an expression for the total number of particles is found by
integrating \eq{Substitution}
\begin{equation}
N_i(t) = N_i(0) -\int_0^t \frac{dt'}{\sqrt{t'}}\, \psi_i(t'),\;\;\;\;\;i=1,\ldots,M.
\end{equation}
$N_i(t),\ldots,N_M(t)$ are found by using the quadrature formula
derived for $I_2^n[\psi]$.

\section{Finding a set of  approximative rate equations from an
expansion of \eq{DefofM} in the variable $s$.} \label{appFirstTauM2}
In this section, details for obtaining a set of a first order rate
equations from \eq{DefofM} will be outlined. The dynamical equations
are found by approximating ${\cal M}$ given in \eq{DetM}. ${\cal M}$ is
approximated by using the expansion stated in \eq{Ordo}. The inverse
Laplace transform of \eq{DefofM} after approximation reads
{\small
  \begin{subequations}
  \label{SecondAttempt}
  \begin{eqnarray}
  \dot{N}_1(t) &= &  \frac{3V_d(R_1)V_{\rm tube}}{3V_d(R_1)
                     +V_{\rm tube}}\frac{D}{\ell^2}\left[
                     \frac{N_2(t)}{V_d(R_2)}-\frac{N_1(t)}{V_d(R_1)}\right]\nonumber\\
               & &   -\delta(t)\frac{V_{\rm tube}}{3V_d(R_1)
                     +V_{\rm tube}}N_{10}\\
  \dot{N}_2(t) &= &  \frac{3V_d(R_2)V_{\rm tube}}{3V_d(R_2)+
                     V_{\rm tube}}\frac{D}{\ell^2}
                     \left[\frac{N_1(t)}{V_d(R_1)}
                     -\frac{N_2(t)}{V_d(R_2)}\right]\nonumber\\
               &  &  -\delta(t)\frac{V_{\rm tube}}{3V_d(R_2)
                     +V_{\rm tube}}N_{20}.
  \end{eqnarray}
  \end{subequations}
}
The decay rate predicted by these equations is given by $\tau_{1,b'}^{-1}=
-\ell^2/q^2_{1,b'}D$ where
\begin{equation}\label{Rate2}
q^2_{1,b'} = -\frac{ V_{\rm  tube}[V_d(R_1) + V_d(R_2)] }
                 { V_d(R_2)[V_d(R_1) + V_{\rm tube}/3] }
\end{equation}
This decay exponent is not adequate. For example, in the case where
all volumes are equal $q^2_{1,b'} = -1.5$ while the exact value is
$q_{\rm exact} = -1.71$.

The rate Eq.~(\ref{SecondAttempt}) can not describe the behavior of
$N_{1,2}(t)$ as $t\rightarrow \infty$ and $t\rightarrow 0$: the
values for $N_{1,2}(\infty)$ are given by
\begin{equation}
\frac{N_1(\infty)}{V_d(R_1)} =  \frac{N_2(\infty)}{V_d(R_1)} =
                                \frac{N_{10} + N_{20}}{V_d(R_1)+V_d(R_2)
                                + \frac{2}{3}V_{\rm tube}}
\end{equation}
and for $N_{1,2}(0^+)$ there is a sudden jump
\begin{equation}
  N_i(0^+) = \frac{N_{i0}}{1+V_{\rm tube}/3V_d(R_i)}, \;\;i = 1,2
\end{equation}
and $N_i(0^+)\neq N_i(0)$. This is not satisfactory and another type
of expansion is needed if correct limits and better decay rates are to
be found.

\section{Rate Equations for the Case studies}\label{CaseRates}
This appendix explains in a more detail how to derive the rate
equations used in the cases studies in \sect{CaseStudies}. The
equations are obtained from \eq{GeneralEq}. Also, \eq{LargeVApprox} is
used to illustrate the impacts on dynamics from changes in the network
structure in a less complicated form.  The ${\cal N}_i(t)$,
$\Delta_{ij}(t)$, $\kappa_{ij}(t)$ and $\sigma_{ij}(t)$ are defined in
\sect{ProblemDef}. The initial distribution is set to be $N_j(0) =
N_{j0}$ where $j=1,\ldots,M$ and $M$ is the total number of nodes in
the system. Note the symmetry relations $a_{ij}=a_{ji}$,
$\ell_{ij}=\ell_{ji}$ and $D_{ij}=D_{ji}$ which implies that
$\kappa_{ij}(t) = \kappa_{ji}(t)$, $\Delta_{ij}(t) = \Delta_{ji}(t)$
and $\sigma_{ij}(t)=\sigma_{ji}(t)$ (see Eq. \ref{GeneralRateCoef}).

\subsection{The Line} \label{LineRate}
The structure of the linear network studied here is shown in
\fig{panel}~(a). Containers $C_1$ and $C_3$ are connected only to the
middle container $C_2$. Therefore, the rate equations describing the
dynamics in each container, $N_1(t)$ and $N_3(t)$ respectively, have
the similar form:
\begin{equation}\begin{array}{lll}\label{LineRate1}
\dot{N}_i(t) &= & \frac{V_{d-1}(a_{2i})}{V_d(R_2)}\int_0^t
                 dt' {\cal N}_2(t')\,\sigma_{2i}(t-t')\\\\

             & & -\frac{V_{d-1}(a_{i2})}{V_d(R_i)}\int_0^t dt'{\cal N}_i(t')
                 \left[\kappa_{i2}(t-t') \right.\\\\

             & &\left. + \Delta_{i2}(t-t')\right], \;\;\;\;\; i=1,3.
\end{array}\end{equation}
The middle container $C_2$ is connected both to container $C_1$ and
$C_3$. The rate equations for $N_2(t)$ reads
\begin{equation}\begin{array}{lll}\label{LineRate2}
\dot{N}_2(t) & = & \frac{V_{d-1}(a_{12})}{V_d(R_1)}\int_0^t dt'
                   {\cal N}_1(t')\,\sigma_{12}(t-t')\\\\

             &  & +\frac{V_{d-1}(a_{32})}{V_d(R_3)}\int_0^t dt'
                  {\cal N}_3(t')\,\sigma_{32}(t-t')\\\\

             &  & -\frac{V_{d-1}(a_{21})}{V_d(R_2)}\int_0^t dt'
                  {\cal N}_2(t') [\kappa_{21}(t-t')+\Delta_{21}(t-t')]\\\\

             &  & -\frac{V_{d-1}(a_{23})}{V_d(R_2)}\int_0^t dt' {\cal N}_2(t')
                  [\kappa_{23}(t-t')+\Delta_{23}(t-t')].
\end{array}\end{equation}
Equations (\ref{LineRate1}) and (\ref{LineRate2}) have a quite
complicated structure and impacts on the dynamics of $N_1(t)$,
$N_2(t)$ and $N_3(t)$ due to changes in the structure (e.g tube
lengths and container volumes) are not easily predicted. Simpler
expressions can be obtained by using \eq{LargeVApprox} which is valid
in the large time limit under the assumption that the number of
particles in the tubes is negligible over time. Applying
\eq{LargeVApprox} leads to
\begin{equation}\begin{array}{lll}
\dot{N}_i(t) &= &\frac{V_{d-1}(a_{2i})}{V_d(R_2)}
\frac{D}{\ell_{2i}}N_2(t) \\\\
&&-\frac{V_{d-1}(a_{i2})}{V_d(R_i)}
\frac{D}{\ell_{i2}}N_i(t),\;\;\;\;\;i = 1,3\\\\

\dot{N}_2(t)&=&\frac{V_{d-1}(a_{12})}{V_d(R_2)}
\frac{D}{\ell_{12}}N_1(t)+\frac{V_{d-1}(a_{32})}{V_d(R_3)}
\frac{D}{\ell_{32}}N_3(t)\\\\

&&\hfill-\frac{D}{V_d(R_2)}
\left[\frac{V_{d-1}(a_{12})}{\ell_{12}}
+\frac{V_{d-1}(a_{23})}{\ell_{23}}\right]N_2(t).


\end{array}\end{equation}

\subsection{The triangle}\label{TriangleRate}
The triangular network studied here is depicted in
\fig{panel}~(b). All the containers $C_1$, $C_2$ and $C_3$ are
connected to each other and therefore the rate equations describing
the time evolution of $N_1(t)$, $N_2(t)$ and $N_3(t)$ have the same
form. Equation (\ref{TriangleRate1}) is the rate equation that governs
the dynamics in container $C_1$. The corresponding dynamical equations
for $N_2(t)$ and $N_3(t)$ are obtained in the same way but are not
written down here
\begin{equation}\begin{array}{lll}\label{TriangleRate1}
\dot{N}_1(t) & = & \frac{V_{d-1}(a_{21})}{V_d(R_2)}\int_0^t
                   dt' {\cal N}_2(t')\,\sigma_{21}(t-t')\\\\

             &  & +\frac{V_{d-1}(a_{31})}{V_d(R_3)}\int_0^t dt'
                  {\cal N}_3(t')\,\sigma_{31}(t-t')\\\\

             &  & -\frac{V_{d-1}(a_{12})}{V_d(R_1)}\int_0^t dt'
                  {\cal N}_1(t')[\kappa_{12}(t-t') + \Delta_{12}(t-t')]\\\\

             &  & -\frac{V_{d-1}(a_{13})}{V_d(R_1)}\int_0^t dt'
          {\cal N}_1(t')[\kappa_{13}(t-t') + \Delta_{13}(t-t')].
\end{array}\end{equation}

The behavior of $N_1(t)$, $N_2(t)$ and $N_3(t)$ are very sensitive
to changes in the network structure. The response from these
changes are not easily predicted by \eq{TriangleRate1}. Instead
\eq{LargeVApprox} can be used to state a simplified version of
\eq{TriangleRate1}. Note that this equation only is valid in the
large time limit if the number of particles in the tubes is small.
Applying \eq{LargeVApprox} to container $C_1$ in the triangular
network gives
\begin{equation}\begin{array}{lll}
  \dot{N}_1(t) & = \hfill
                   & \frac{V_{d-1}(a_{21})}{V_d(R_2)}\frac{D}{\ell_{21}}N_2(t) +
                   \frac{V_{d-1}(a_{31})}{V_d(R_3)}\frac{D}{\ell_{31}}N_3(t)\\\\
               &&  \hfill -\frac{D}{V_d(R_1)}\left[\frac{V_{d-1}(a_{12})}{\ell_{12}}+
                   \frac{V_{d-1}(a_{13})}{\ell_{13}}\right]N_1(t).
  \end{array}\end{equation}
Corresponding first order rate equations for $N_2(t)$ and $N_3(t)$ are
found in the same way.
%
%

\subsection{The T-Network}\label{Trate}
The T-shaped network under investigation in this subsection is
depicted in \fig{panel}~(c). Since all containers $C_1$, $C_2$ and
$C_3$ are connected to container $C_4$ and not to each other, the rate
equations governing the dynamics in $C_1$, $C_2$ and $C_3$, that is
$N_1(t)$, $N_2(t)$ and $N_3(t)$ respectively, all have similar form,
\begin{equation}\begin{array}{lll}\label{JunctionRate11}
\dot{N}_i(t) & = & \frac{V_{d-1}(a_{4i})}{V_d(R_4)}\int_0^t dt'
                   {\cal N}_4(t')\,\sigma_{4i}(t-t')\\\\

             &  & -\frac{V_{d-1}(a_{i4})}{V_d(R_i)}\int_0^t dt'
           {\cal N}_i(t') \left[\kappa_{i4}(t-t') \right.\\\\

         &  & \left. + \Delta_{i4}(t-t')\right], \;\;\;\;\; i=1,2,3.
\end{array}\end{equation}
%
%
%
%
The middle container $C_4$ has connections to all other containers
$C_1$, $C_2$ and $C_3$. This leads to a rate equation for $N_4(t)$
on the form
\begin{equation}\begin{array}{lll}\label{JunctionRate14}
\dot{N}_4(t) &= & \sum_{j=1}^3 \frac{V_{d-1}(a_{j4})}{V_d(R_j)}
                  \int_0^t dt' {\cal N}_j(t')\,\sigma_{j4}(t-t')\\\\
             &  & -\frac{1}{V_d(R_4)}\int_0^t dt' {\cal N}_4(t')\\\\
             &  & \times \sum_{j=1}^3 V_d(a_{4j})[\kappa_{4j}(t-t') +
          \Delta_{4j}(t-t')].
\end{array}\end{equation}
The rate equations derived in this subsection are used in the study of
the junction, shown in \fig{Junction1}.

\section{Elimination of junctions in the asymptotic regime}\label{SectJunction}
The network studied in this section is a generalization of the one
depicted in \fig{Junction1}. Here, a junction having an arbitrary
number of connections is studied and it will be demonstrated that the
dynamical equations governing the transport in a system having
junction points can be simplified in the large time limit. First, it
will be shown that the current of particles in and out of a junction
point eventually will balance each other out and that this occurs
faster than the time it takes before an equilibrium particle
distribution is attained throughout the network. This derives from the
fact that the volume of the junction \mbox{(proportional to $ a^d$)}
is much smaller than the volumes of the containers,
\mbox{(proportional to $R_i^d$, $i=1,\ldots,M$)}, where $d$ is the
dimensionality, $a$ is the tube radius and $M$ is the number of
containers. The tubes connecting the junction are assumed to have the
same radius. Second, it will be demonstrated how to simplify the
transport equations in such a way that the dynamical variable for the
junction $N(t)$ can be eliminated completely. This might be desirable
when working with large networks.

Consider a junction with dynamical behavior contained in $N_j(t)$ and
volume $V_d(a)$ that has connections to $M$ containers with volumes
$V_d(R_i)$, $i=1,\ldots,M$. The equation governing the dynamics of the
junction point written in Laplace space reads
\begin{equation}\begin{array}{lll}\label{JuncRateInS}
sN_j(s)-N_j(0) &= & -sN_j(s) \frac{V_{d-1}(a)}{V_d(a)}\sum_{i=1}^M\left[
                    \kappa_{ji}(s) + \Delta_{ji}(s)\right]\\\\
               & &  +\sum_{i=1}^M sN_i(s)\frac{V_{d-1}(a)}{V_d(R_i)}
                    \,\sigma_{ij}(s).
\end{array}\end{equation}
This equation is a generalization of \eqs{RateInS} (see
\sect{SecGeneralEqs}). The junction point studied here is such that $a
\ll R_i$ and since
\begin{equation}\begin{array}{lr}
\frac{V_{d-1}(a)}{V_d(a)}= c_d\frac{1}{a},&
\;\;\;\;\;\;\frac{V_{d-1}(a)}{V_d(R_i)} =
c_d\frac{1}{a}\left(\frac{a}{R_i}\right)^d
\end{array}\end{equation}
where $c_d = \frac{d}{d-1}\Gamma\left(\frac{d}{2}\right)
[\Gamma\left(\frac{d-1}{2}\right)]^{-1}\,\pi^{-1/2}$ [found by using
the formula for a $d$ - dimensional sphere, see \sect{ProblemDef}] the
term proportional $(a/R_i)^d$ can be neglected and a closed expression
for $N_j(s)$ can be obtained. Solving \eq{JuncRateInS} after eliminating
the second term leads to
\begin{equation}\label{JuncSol}
N_j(s) = \frac{N_j(0)}{s\left(1 + \frac{c_d}{a}\sum_{i=1}^M\left[
\kappa_{ji}(s) + \Delta_{ji}(s)\right] \right)}.
\end{equation}
The inverse Laplace transform of $N_j(s)$ is a sum of exponentials
\begin{equation}\label{SumOfExp}
N_j(t) = \sum_{p=1}^\infty \alpha_p e^{-\beta_pt}
\end{equation}
where $\alpha_p$ is the residue of $N_j(s)$ at pole $s = -\beta_p$,
\mbox{$\beta_p\in \Re>0$}. The poles are given by the zeros of
\begin{equation}\label{TransEqJunction}
s+c_d\sum_{i=1}^M\frac{q_{ij}D_{ij}}{a\ell_{ij}} \frac{\cosh
q_{ji}} {\sinh q_{ij}}=0
\end{equation}
where $q_{ij}^2=s\ell_{ij}^2/D_{ij}$. When the currents in and out of
the junction point balances each other out there is no accumulation of
particles and $\dot{N}_j(t)=0$.  This is easily verified from
\eq{SumOfExp} by evaluating the derivative in respect to $t$ at
$t=\infty$
\begin{equation}
\frac{d}{dt} N_j(t) = \sum_{p=1}^\infty (-\beta_p) \alpha_p
e^{-\beta_pt} \rightarrow 0 \;\;{\rm as } \;\; t\rightarrow \infty.
\end{equation}
Let $\tau_{\rm junction}$ be an estimate of the time it takes until
$\dot{N}_j(t)=0$. It is related to the decay exponents $\beta_p$ (see
Eqs. \ref{tau_s1} and \ref{q_s1}) which are zeros of
\eq{TransEqJunction}. From \eq{TransEqJunction} it is clear that the
zeros scales with $\frac{D}{a\ell}$ (it is assumed that $D_{ij}\sim D$
and $\ell_{ij} \sim \ell$) which leads to the estimate $\tau_{\rm
junction} \sim \frac{a\ell}{D}$. Let $\tau_{\rm network}$ be an
estimate of the time it takes to reach an overall equilibrium particle
distribution in the network.  As a rough estimate \eq{Rate1} can be
used which was found for a two-node network. If the containers have
volumes proportional to $R^d$ then $\tau_{\rm network} \sim
\frac{a\ell}{D}\left(\frac{R}{a}\right)^d$. If $R\gg a$ then
$\tau_{\rm junction}\ll\tau_{\rm network}$. This is is supported by
the numerical calculation shown in \fig{Junction} where the curve
corresponding to $N_4(t)$ flattens out relatively fast.

For networks where there are many junctions and containers involved it
might be desirable to reduce the number of variables in the dynamical
equations governing the particle transport. This can be done in the
regime where $\dot{N}_j(t) = 0$ is valid and \eq{LargeVApprox} is
applicable. Applying \eq{LargeVApprox} for a junction point $j$ having
$M$ connections and using $\dot{N}_j(t)=0$ leads to
\begin{equation}
0 = \sum_{i=1}^M {\cal C}_{ij}V_{d-1}(a)\frac{D_{ij}}{\ell_{ij}}
\left[\frac{N_i(t)}{V_d(R_i)}-\frac{N_j(t)}{V_d(a)}\right]
\end{equation}
In this way it is possible to express $N_j(t)$ in terms of
$N_1(t),\ldots,N_M(t)$. Inserting the solution to this matrix equation
in the dynamical equations eliminates the explicit dependence of
$N_j(t)$. \eqs{RateEqsJunction21} - (\ref{RateEqsJunction23}) shows
this explicitly for the case of a three way junction depicted in
\fig{Junction1}, where $N_4(t)$ has been taken away completely
\begin{equation}\begin{array}{lll}\label{RateEqsJunction21}
\dot{N}_1(t) &= & \frac{\ell_{24} D}{\ell^2}
                  \frac{V_{d-1}(a)}{V_d(R_2)}N_2(t) +\frac{\ell_{34}D}{\ell^2}
                  \frac{V_{d-1}(a)}{V_d(R_3)}N_3(t)\\\\
             & &  -\frac{(\ell_{24}+\ell_{34})D}{\ell^2}
                  \frac{V_{d-1}(a)}{V_d(R_1)}N_1(t)
\end{array}\end{equation}
\begin{equation}\begin{array}{lll}\label{RateEqsJunction22}
\dot{N}_2(t) & = & \frac{\ell_{14}D}{\ell^2}
                   \frac{V_{d-1}(a)}{V_d(R_1)}N_1(t)
                   +\frac{\ell_{34}D}{\ell^2}\frac{V_{d-1}(a)}{V_d(R_3)}
                   N_3(t)\\\\
             &  &  -\frac{(\ell_{14}+\ell_{34})D}
                   {\ell^2}\frac{V_{d-1}(a)} {V_d(R_2)}N_2(t)\\\\\\
\end{array}\end{equation}
\begin{equation}\begin{array}{lll}\label{RateEqsJunction23}
\dot{N}_3(t) &= & \frac{\ell_{14}D}{\ell^2}
                  \frac{V_{d-1}(a)}{V_d(R_2)}N_1(t)
                  +\frac{\ell_{24}} {\ell^2}
                  \frac{V_{d-1}(a)}{V_d(R_2)}N_2(t)\\\\
             & &  -\frac{(\ell_{14}+\ell_{24})D}{\ell^2}
                  \frac{V_{d-1}(a)}{V_d(R_3)}N_3(t)\\\\
                  \ell^2 &\equiv& \ell_{14}\ell_{24} +
          \ell_{24}\ell_{34} + \ell_{14}\ell_{34}.
\end{array}\end{equation}

\vfill



%
\newpage

\begin{figure}
\centering
\psfrag{Ri}[t][t]{\footnotesize{$R_i$}}
\psfrag{a}[t][t]{\footnotesize{$2a_{ij}$}}
\includegraphics[width = 7cm]{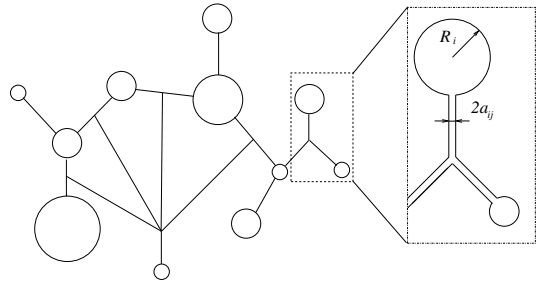}
\caption{ Schematic picture of an arbitrary network built from
containers and tubes.}
\label{MainNetwork}
\end{figure}

\begin{figure}
\psfrag{x = 0}[][]{$x=0$}
\psfrag{R}[][]{$R$}
\psfrag{2a}[][]{$2a$}
\psfrag{1}[][]{$(a)$}
\psfrag{2}[][]{$(b)$}
\centering
\includegraphics[width = 7cm]{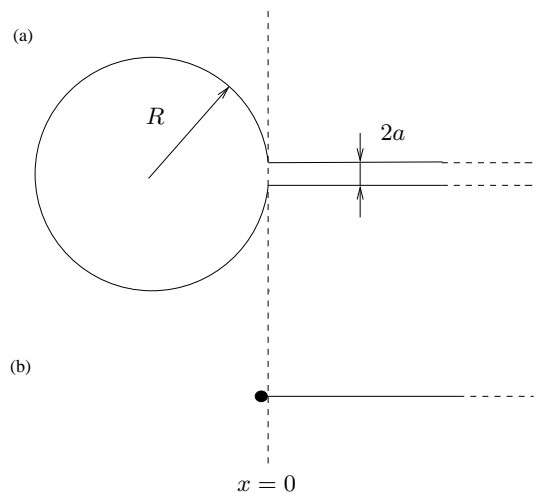}
\caption{Panel (a): A spherical compartment connected to a
cylindrical infinitely long tube. If the tube radius is assumed to be
small the compartment can be treated as ideally mixed at all times
simplifying the dynamics in the container. The transport in the tube
is reduced to a one dimensional diffusion problem. Panel (b)
illustrates these simplifications.}
\label{InfTubeFig}
\end{figure}

\begin{figure}
\psfrag{t1}[][]{{\small $\frac{Dt_1}{\ell^2} = 0.001$}}
\psfrag{t2}[][]{{\small $\frac{Dt_2}{\ell^2} = 0.01$}}
\psfrag{t3}[][]{{\small $\frac{Dt_3}{\ell^2} = 1$}}
\psfrag{xlabel}[][]{$y/2a$}
\psfrag{ylabel}[][]{$2a f(y,t)$}
\centering
\includegraphics[width = 8cm]{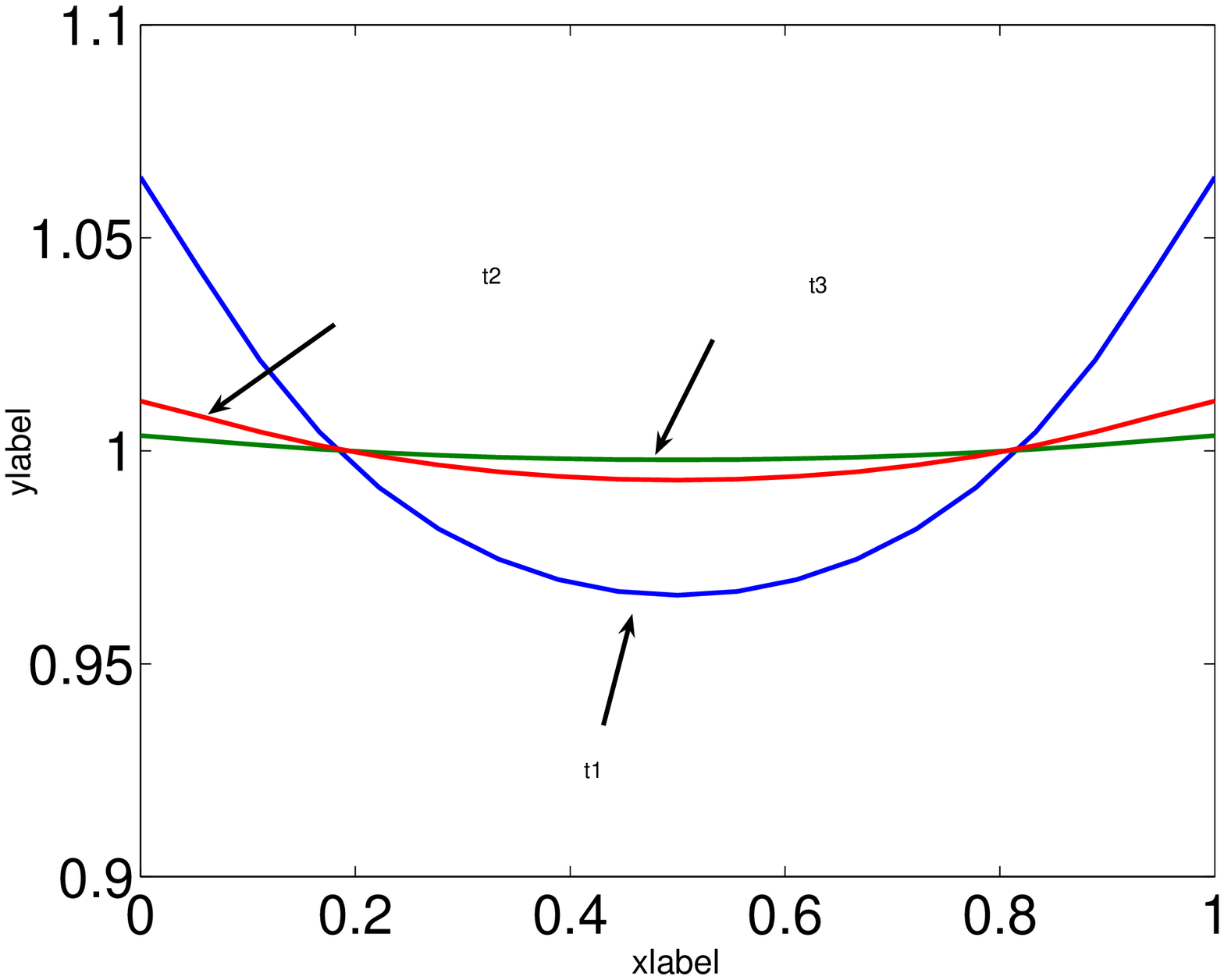}
\caption{Behavior of $f(y,z,t)$ defined in
\eq{DeCoupling} for the two dimensional system shown in
\fig{SolDiffEq} [panel~(a), $a/R = 0.1$] at three different instants of
time $t_1<t_2<t_3$. The graph verifies the assumption that
$f(y,z,t)$ is approximately constant. In this two dimensional case
$f(y,t) \rightarrow 1/2a$ as $t \rightarrow \infty$.}
\label{InletProfile}
\end{figure}

\begin{figure}
\psfrag{x1}[][]{$x=0$}
\psfrag{x2}[][]{$x=\ell$}
\psfrag{a}[][]{$2a$}
\psfrag{R1}[][]{$R_1$}
\psfrag{R2}[][]{$R_2$}
\centering
\includegraphics[width = 7cm]{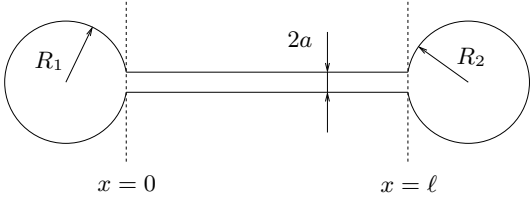}
\caption{Schematic picture of a two-node network.}
\label{TwoContSytem}
\end{figure}

\begin{figure}
\psfrag{xlabel}[][]{$Dt/\ell^2$}
\psfrag{ylabel}[][]{$\displaystyle \frac{\ell}{D}\left[\Delta(t),
\kappa(t),\sigma(t)\right]$}
\centering
\includegraphics[width = 8cm]{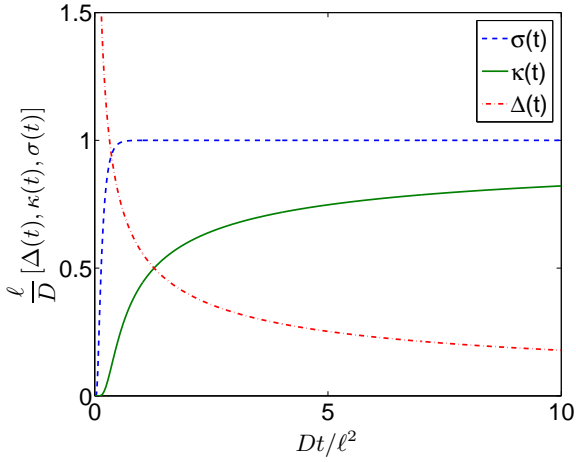}
\caption{Time dependence of the rate coefficients $\Delta(t)$,
$\kappa(t)$ and $\sigma(t)$ from \eq{GeneralEq}.}
\label{RateGraph}
\end{figure}

\begin{figure}
\psfrag{xlabel}[][]{$Dt/\ell^2$}
\psfrag{ylabel}[][]{$N_1(t)/N_{tot}$}
\psfrag{V1}[][]{$N_1, V_1$}
\psfrag{V2}[][]{$N_2, V_2$}
\psfrag{LL}[][]{$\ell$}
\psfrag{11}[][]{\small{$C_1$}}
\psfrag{22}[][]{\small{$C_2$}}
\centering
\includegraphics[width = 8cm]{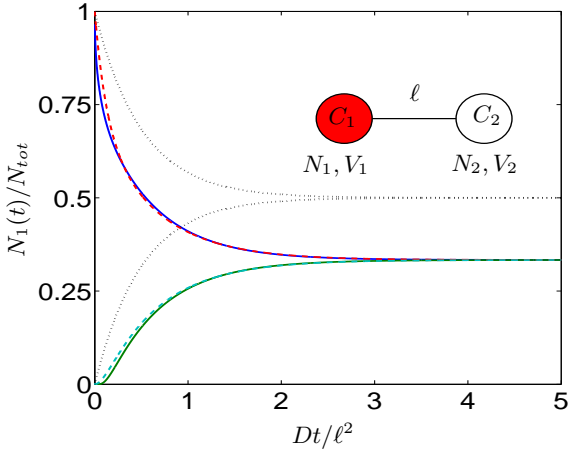}
\caption{The solution of \eq{RateEqTwoCont} (solid line) is compared
with a solution of \eq{ThirdAttempt} (dashed line).  The solution of
\eq{DagdugRateEqn}, given in (\ref{DagdugApprox}), is represented by
the dotted line. The curves decaying and growing describe $N_1(t)$ and
$N_2(t)$ respectively. The network structure is shown in inset. The
volumes of the tube and the containers are equal and $a/\ell =
0.05$. The initial distribution of particles is $N_1(0)/N_{tot} = 1$
and $N_2(0)/N_{tot}=0$, where $N_{tot}=N_1(0) + N_2(0)$, indicated by
the shading in inset. This figure clearly shows that \eq{DagdugApprox}
does not lead to the correct values for $N_1(t)$ and $N_2(t)$.  The
agreement with \eq{ThirdAttempt} is much better.}
\label{NumSolTwoCont}
\end{figure}

\begin{figure}
\psfrag{xlabel}[][]{$Dt/\ell^2$}
\psfrag{ylabel}[][]{$\displaystyle \log \left[\frac{N_1(t)-N_1(\infty)}
{N_{tot}}\right]$}
\centering
\includegraphics[width = 8cm]{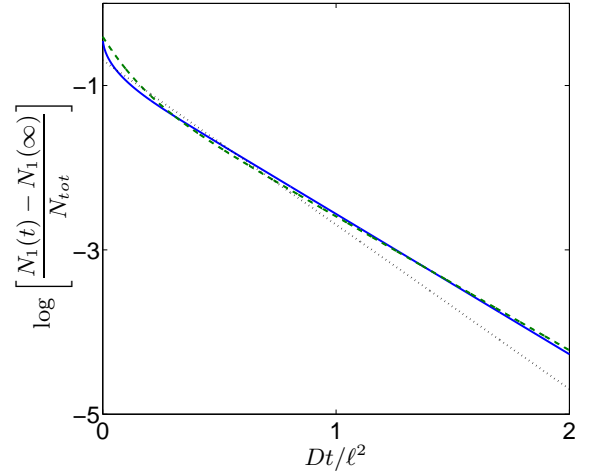}
\caption{The natural logarithm of $[N_1(t)-N_1(\infty)]/N_{tot}$ for
the three cases illustrated in \fig{NumSolTwoCont}. The labeling of
the curves is the same as in \fig{NumSolTwoCont}. The linear behavior
is the evidence of the single exponential decay of the number of
particles in the container $C_1$. The slope gives the value of the decay
exponent. The slopes of the solid and dashed lines are close to each
other showing that \eq{ThirdAttempt} is capable of estimating the
decay exponent well. The slope of the dotted line differs
significantly from the others which illustrates that \eq{DagdugApprox}
can not describe the dynamics in an adequate way.  Since the value of
$N_1(\infty)$ is not the same as in all three cases [compare
\eq{Ninfty} and (\ref{LimitDist})] the value of $N_1(0) - N_1(\infty)$
will be different. This explains why all three curves do not coincide
at $t=0$.}
\label{LogTwoCont}
\end{figure}

\begin{figure}
\psfrag{xlabel}[][]{$V_{\rm tube}/V_d$}
\psfrag{ylabel}[][]{{\large $q_1^2$}}
\centering
\includegraphics[width = 8cm]{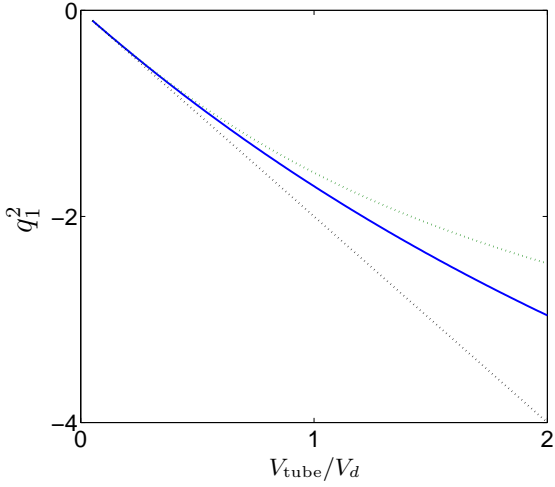}
\caption{Dependence of the geometrical factor $q^2$ on the tube
volume. (The volumes of the containers are equal
$V_d(R_1)=V_d(R_2)\equiv V_d$.)  $q_1^2$ and $\tau^{-1}$ are related
through $\tau^{-1}= -q_1^2D/\ell^2$. The numerical solution of
\eq{TransEq}, which gives the exact values for $q_1^2$, is represented
by the solid line. It is compared with the values for $q_{1,a}^2$
(Eq. \ref{Rate1}), dashed line, and $q_{1,b}^2$ (Eq. \ref{Rate3}),
dotted line. The dotted line deviates significantly from the solid one
as $V_{\rm tube}/V_d$ increases while the dashed line follows the
exact solution better. This indicates that $q_{1,b}^2$ provides a good
estimate of the decay rate, even for large tube volumes. $q_{1,a}^2$
can only be used for very small values of $V_{\rm tube}/V_d$.}
\label{TransEqFig}
\end{figure}

\begin{figure}
\psfrag{XX}[][]{$Dt/\ell^2$}
\psfrag{YY}[][]
{$\displaystyle \log \left[ \frac{N_1(t)-N_1(\infty)}{N_{tot}}\right]$}
\psfrag{2a}[][]{\footnotesize{$2a$}}
\psfrag{L}[][]{\footnotesize{$2R$}}
\psfrag{l}[][]{\footnotesize{$\ell$}}
\centering
\includegraphics[width = 8cm]{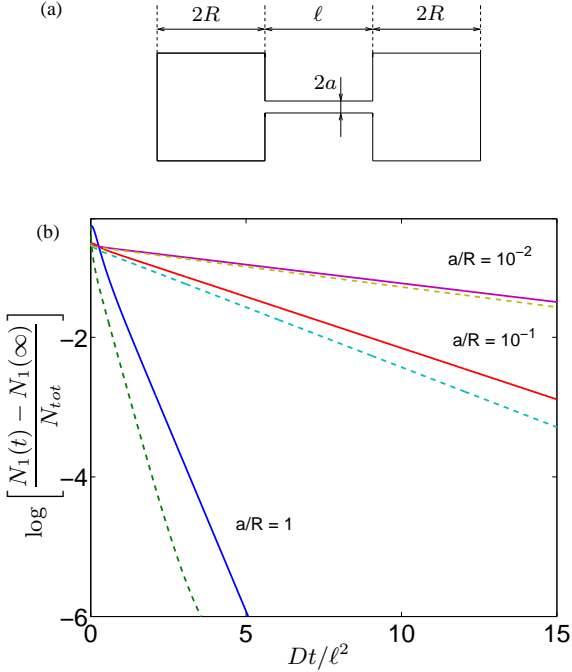}
\caption{A numerical verification of the assumption of well stirred
containers. For simplicity reasons a cubic geometry, as depicted in
panel (a), was chosen since the explicit shape of the container looses
its importance for large reservoir volumes. The edge length is set to
$2R$ where $R$ is the radius of an equivalent spheric container.
Panel (b) shows a numerical solution to the 2D-diffusion equation
(solid line) compared to a numerical solution to \eqs{GeneralEq}
(dashed line). The cases are chosen so that $a/R$ varies in three
orders of magnitude showing increasing validity of the assumption of
ideally mixed containers as $a/R$ decreases. The initial distribution
in all three cases are skewed (a delta function in the bottom right
corner of the left container) which becomes important when $a$ is
large. For small $a$, $\tau_{{\rm mix}}$ is a lot shorter than
$\tau_{{\rm target}}$ and the skewed initial distribution will have
time to smear out before particles start exiting and the shape of the
initial distribution has no effect.}
\label{SolDiffEq}
\end{figure}

\begin{figure}
\psfrag{L12}[][]{\footnotesize{$\ell_{12}$}}
\psfrag{L23}[][]{\footnotesize{$\ell_{23}$}}
\psfrag{L24}[][]{\footnotesize{$\ell_{24}$}}
\psfrag{L31}[][]{\footnotesize{$\ell_{31}$}}
\psfrag{L14}[][]{\footnotesize{$\ell_{14}$}}
\psfrag{L34}[][]{\footnotesize{$\ell_{34}$}}
\psfrag{N1}[][]{\small{$N_1$, $R_1$}}
\psfrag{N2}[][]{\small{$N_2$,$R_2$}}
\psfrag{N3}[][]{\small{$N_3$, $R_3$}}
\psfrag{N4}[][]{\small{$N_4$, $R_4$}}
\psfrag{1}[][]{\small{$C_1$}}
\psfrag{2}[][]{\small{$C_2$}}
\psfrag{3}[][]{\small{$C_3$}}
\psfrag{4}[][]{\small{$C_4$}}
\includegraphics[width = 8cm]{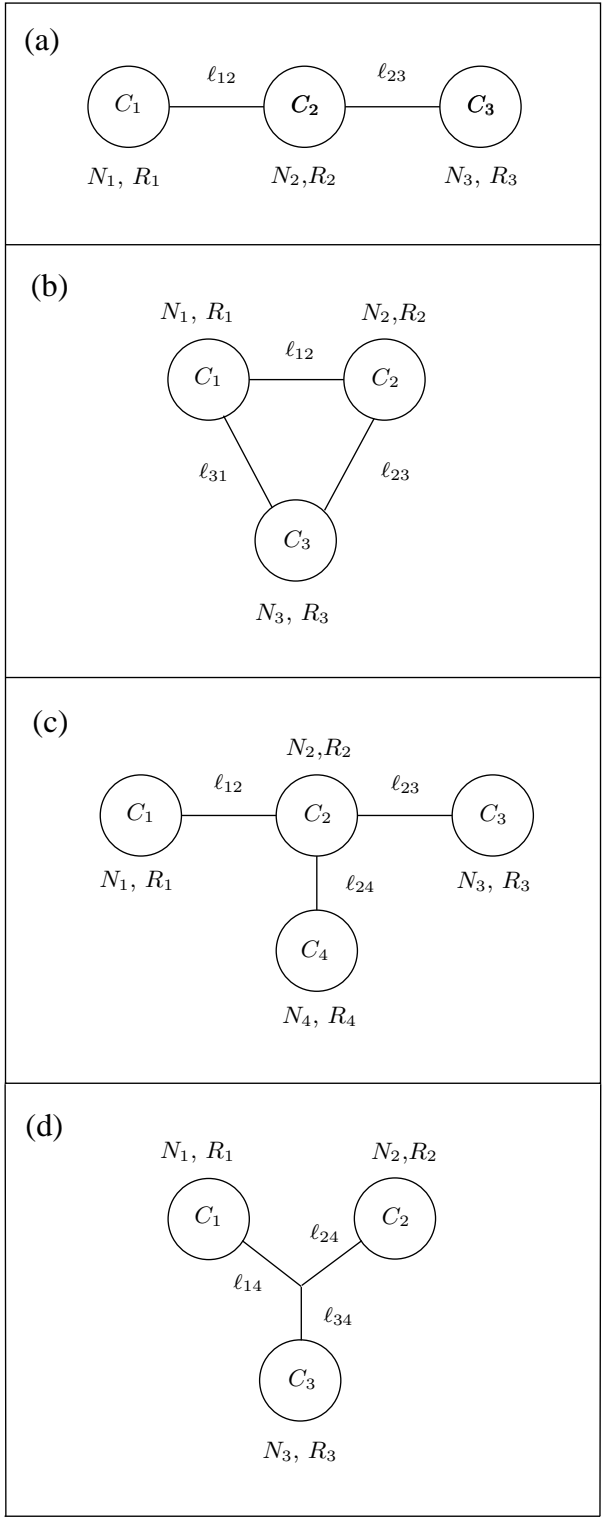}
\caption{networks used for case studies in
\sect{CaseStudies}. Panels~(a)-(d) corresponds to cases 1-4.}
\label{panel}
\end{figure}

\begin{figure}[!ht]
\centering
\psfrag{XX}[][]{\footnotesize{$Dt/\ell^2$}}
\psfrag{YY}[][]{\footnotesize{$N_i(t)/N_{tot}$}}
\psfrag{A}[][]{$a)$}
\psfrag{B}[][]{$b)$}
\psfrag{N1}[][]{\small{$N_1$, $V_1$}}
\psfrag{N2}[][]{\small{$N_2$, $V_2$}}
\psfrag{N3}[][]{\small{$N_3$, $V_3$}}
\psfrag{L12}[][]{\small{$\ell_{12}$}}
\psfrag{L23}[][]{\small{$\ell_{23}$}}
\psfrag{L31}[][]{\small{$\ell_{31}$}}
\includegraphics[width = 8cm]{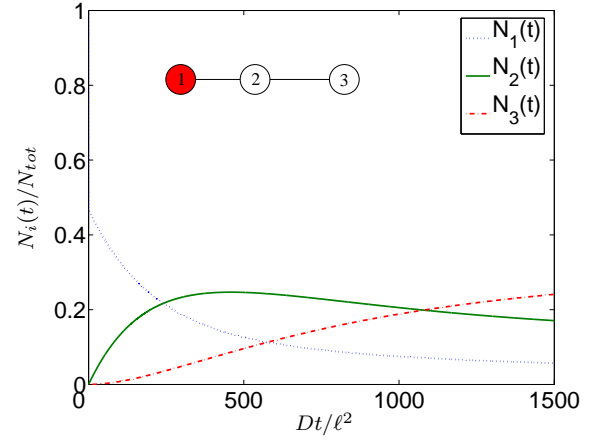}
\caption{The transport properties of the structure depicted in
\fig{panel}~(a), see \sect{CaseStudies} (case 1) for discussion.  The
curves depict a numerical solution of Eqs. (\ref{LineRate1}) and
(\ref{LineRate2}) given in \app{LineRate}. The curve depicting the
number of particles in the middle container $C_2$ (solid line) has a
maximum. This kind of behavior does not exist for the two-node network
where there is only exponential growth and decay (see
\fig{NumSolTwoCont}). The network parameters were set to
$\ell_{12}=\ell_{23}\equiv \ell$, $a/\ell= 1/5$, $a/R_1 = 1/10$,
$a/R_1 = 1/15$, $a/R_3 = 1/20$. The initial distribution of particles
is $N_1(0)/N_{tot} = 1$, $N_2(0)/N_{tot} = N_3(0)/N_{tot} = 0$, shown
graphically in inset.}
\label{LineGraph}
\end{figure}

\begin{figure}[!ht]
\centering
\psfrag{xlabel}[][]{\footnotesize{$Dt/\ell^2$}}
\psfrag{ylabel}[][]{\footnotesize{$N_i(t)/N_{tot}$}}
\includegraphics[width = 8cm]{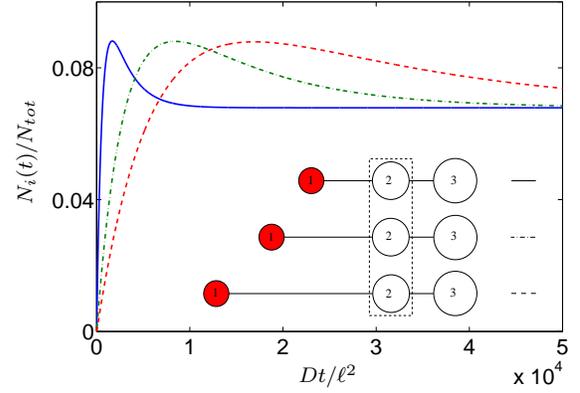}
\caption{The control of the particle arrival time for three different
cases shown in inset: $\ell=\ell_1$ (solid line), $\ell/\ell_2 = 1/5$
(dash-dotted line) and $\ell/\ell_3 = 1/10$ (dashed line). The system
is otherwise equivalent to the one studied in \fig{LineGraph}.  The
initial distribution of particles is shown graphically in inset.}
\label{LineTiming}
\end{figure}

\begin{figure}[!ht]
\centering
\psfrag{A}[][]{$a)$}
\psfrag{B}[][]{$b)$}
\psfrag{xlabel}[][]{\footnotesize{$Dt/\ell^2$}}
\psfrag{ylabel}[][]{\footnotesize{$N_i(t)/N_{tot}$}}
\includegraphics[width = 8cm]{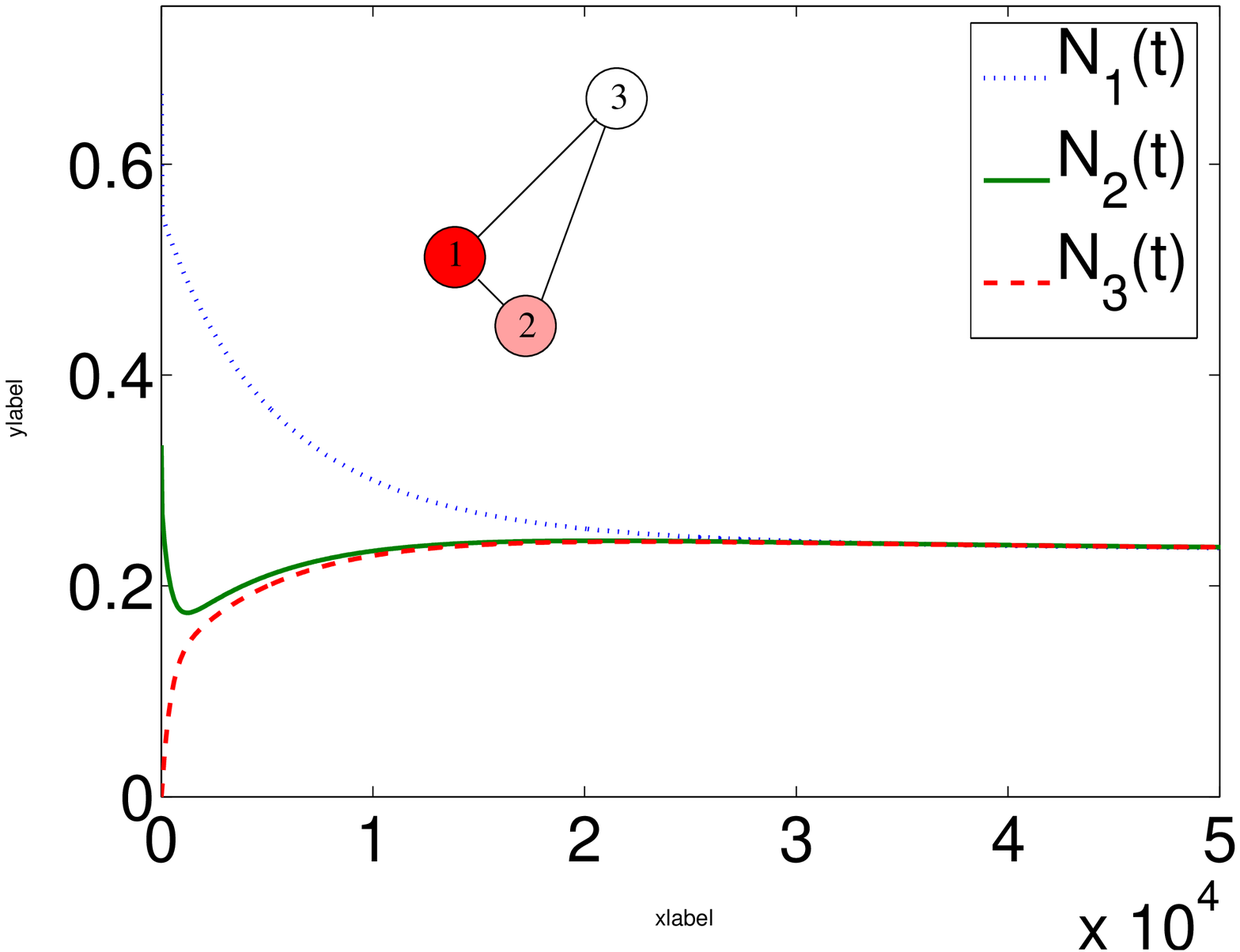}
\caption{The transport properties for the triangular network shown in
\fig{panel} (b), see \sect{CaseStudies} (case 2) for discussion.  The
curves depict a numerical solution of \eq{TriangleRate1} given in
\app{TriangleRate}. The curve depicting the number of particles in
$C_2$ (solid line) has a minimum. This does not occur in the transport
dynamics for the two-node network where there is only exponential
growth and decay (see \fig{NumSolTwoCont}). The parameters were set to
$\ell\equiv\ell_{21}$, $\ell/a=1$, $\ell_{23}/\ell=1/10$,
$\ell_{31}/\ell=50$, $R_1/\ell = R_2/\ell = R_3/\ell=4$. The initial
distribution of particles is $N_1(0)/N_{tot}=1$, $N_2(0)/N_{tot} = 0.5
$, $N_3(0)/N_{tot}=0$ as indicated by shading in inset.}
\label{Triangle2}
\end{figure}

\begin{figure}[!ht]
\centering \psfrag{XX}[][]{$Dt/\ell^2$}
\psfrag{YY}[][]{$N_i(t)/N_{tot}$}
\includegraphics[width = 8cm]{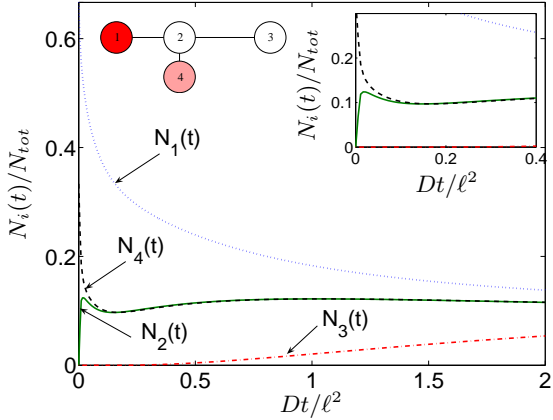}
\caption{The transport properties for the T-shaped network shown in
\fig{panel} (c), see \sect{CaseStudies} (case 3) for discussion.  The
curves depict a numerical solution of \eqs{JunctionRate11} and
(\ref{JunctionRate14}) given in \app{TriangleRate}. The figure shows
two extremum points in the curve depicting the number of particles in
container $C_2$ (solid line). It is not possible to have this kind of
behavior for any of the cases studied involving two and three
containers. The additional extremum point was produced by adding an
additional container $C_4$ to the middle container $C_2$ in the linear
structure shown in \fig{panel}~(a). The network parameters used were
$\ell_{12}\equiv \ell$, $\ell/a = 3$, $\ell/\ell_{23}=1/2$,
$\ell/\ell_{24}=10$, $R_1/\ell=2$. The initial distribution was set to
$N_1(0)/N_{tot}=2/3$, $N_2(0)/N_{tot} = 0$, $N_3(0)/N_{tot} = 0$ and
$N_4(0)/N_{tot} = 1/3$ according to shading in inset.}
\label{LineGraphTop}
\end{figure}

\begin{figure}
\centering
\psfrag{N1}[][]{\small{$N_1$, $V_1$}}
\psfrag{N2}[][]{\small{$N_2$, $V_2$}}
\psfrag{N3}[][]{\small{$N_3$, $V_3$}}
\psfrag{N41}[][]{\small{$N_4$, $V_4$}}
\psfrag{N42}[][]{\small{$N_4$, $V(a)$}}
\psfrag{L14}[][]{\small{$\ell_{14}$}}
\psfrag{L24}[][]{\small{$\ell_{24}$}}
\psfrag{L34}[][]{\small{$\ell_{34}$}}
\psfrag{limit}[][]{\small{$R_4\longrightarrow a$}}
\includegraphics[width = 8cm]{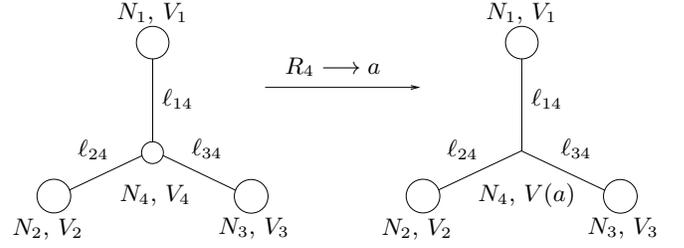}
\caption{The transformation from a four-node network to a network
involving a three way junction.}
\label{Junction1}
\end{figure}

\begin{figure}[!ht]
\centering
\psfrag{A}[][]{$a)$}
\psfrag{B}[][]{$b)$}
\psfrag{xlabel}[][]{\footnotesize{$Dt/\ell^2$}}
\psfrag{ylabel}[][]{\footnotesize{$N_i(t)/N_{tot}$}}
\includegraphics[width = 8cm]{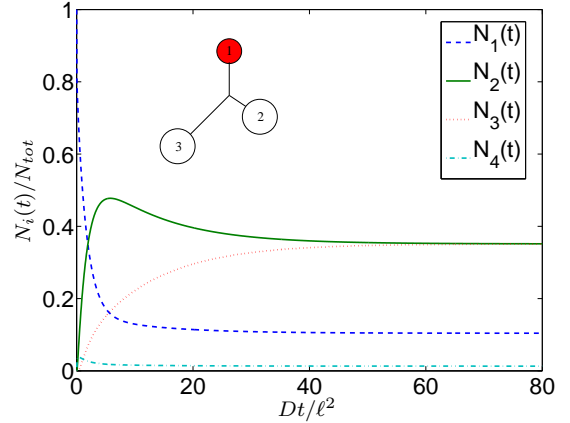}
\caption{The transport properties for the network involving a three
way junction shown in \fig{panel} (d), see \sect{CaseStudies} (case 4)
for discussion.  The curves depict a numerical solution of
\eqs{JunctionRate11} and (\ref{JunctionRate14}) [with $V_d(R_4) =
V_d(a)$]. The system parameters were set to $\ell_{12}/a=2$,
$\ell_{13}/a=10$, $\ell_{14}/a=20$, $R_1/a = 2$ and $R_2/a = R_3/a =
4$. Initial distribution of particles $N_1(0)/N_{tot} = 1$,
$N_2(0)/N_{tot} = N_3(0)/N_{tot}=N_4(0)/N_{tot}=0$ is shown in inset.}
\label{Junction}
\end{figure}

\begin{figure}[!ht]
\centering
\psfrag{1}[][]{\small{$C_1$}} \psfrag{2}[][]{\small{$C_2$}}
\psfrag{3}[][]{\small{$C_3$}} \psfrag{4}[][]{\small{$C_4$}}
\psfrag{5}[][]{\small{$C_5$}} \psfrag{6}[][]{\small{$C_6$}}
\psfrag{7}[][]{\small{$C_7$}} \psfrag{8}[][]{\small{$C_8$}}
\includegraphics[width = 7cm]{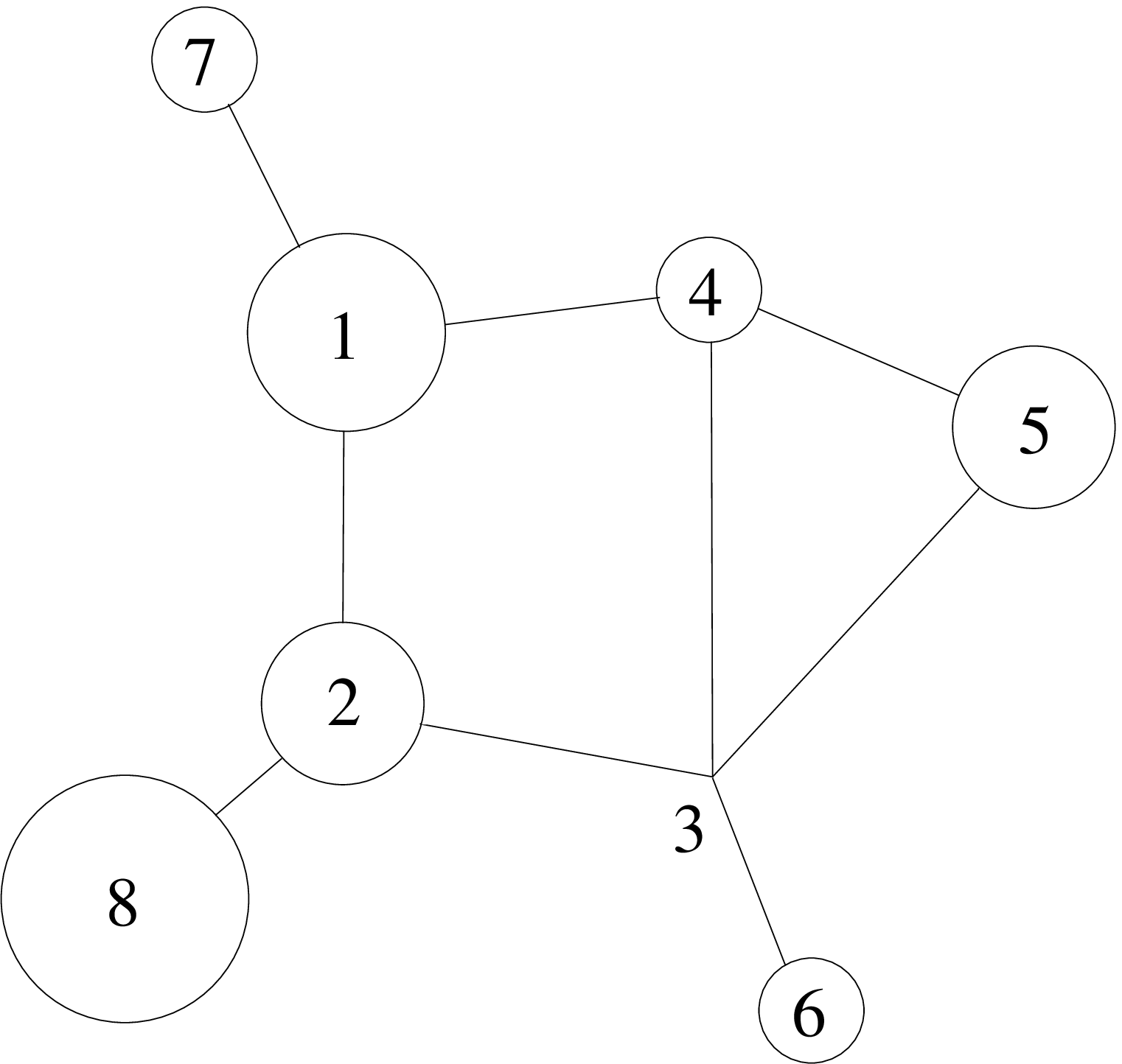}
\caption{Structure of the network studied in \sect{CaseStudies},
case~5. The parameters describing the geometry are labeled in the
same way as in \fig{panel}. The parameters used were $a/\ell=1/3$,
$\ell_{23}/\ell=1$, $\ell_{45}/\ell=9$, $\ell_{43}/\ell=10$,
$\ell_{35}/\ell=8$, $\ell_{36}/\ell=2$, $\ell_{28}/\ell=0.1$,
$\ell_{12}/\ell=1$, $\ell_{32}/\ell=1$, $R_1/a = 4$, $R_2/a = 2$,
$R_3/a = 1$, $R_4/a = 2.5$, $R_5/a = 3$, $R_6/a = 2$, $R_7/a = 2$,
$R_8/a = 4$.}
\label{LargeNetwork}
\end{figure}

\begin{figure}[!ht]
\centering
\psfrag{XX}[][]{\footnotesize{$Dt/\ell^2$}}
\psfrag{YY}[][]{\footnotesize{$N_i(t)/N_{tot}$}}
\includegraphics[width = 8cm]{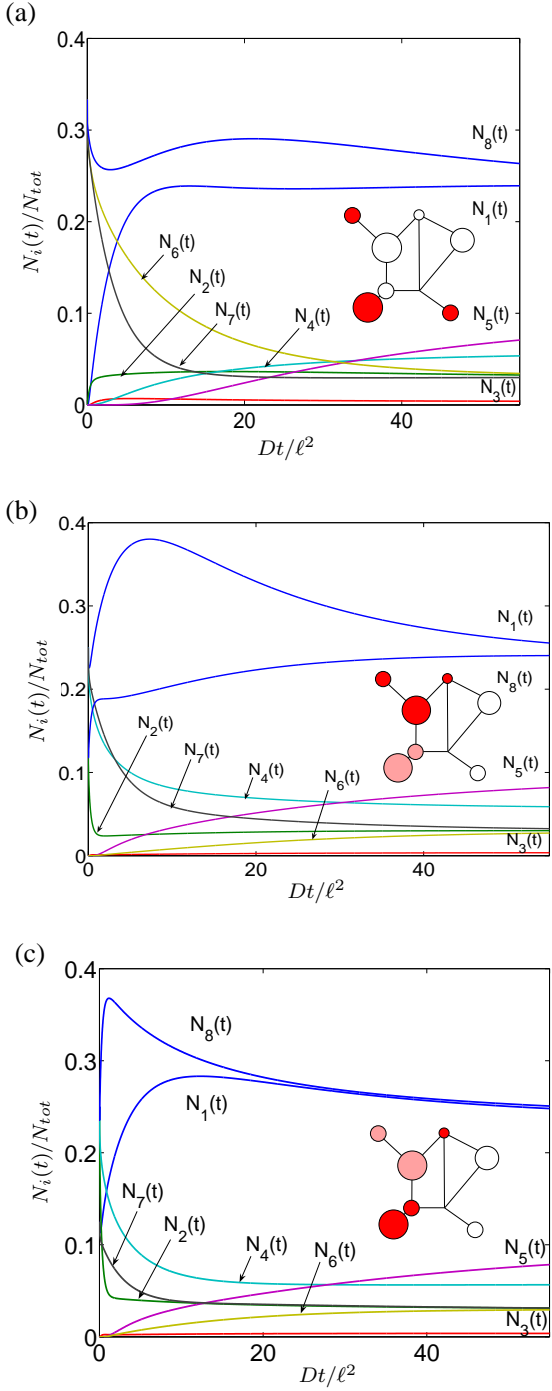}
\caption{The solution of \eq{GeneralEq} for the network depicted in
\fig{LargeNetwork}.  The panel includes graphs showing the transport
dynamics for three different choices of initial distribution of
particles. Panel~(a): $\tilde{N}_1(0) = \tilde{N}_2(0) =
\tilde{N}_3(0) = \tilde{N}_4(0)= \tilde{N}_5(0) =0$, $\tilde{N}_6(0) =
\tilde{N}_7(0) = \tilde{N}_8(0) = 1/3$; Panel~(b): $\tilde{N}_3(0) =
\tilde{N}_5(0) = \tilde{N}_6(0)=0$, $\tilde{N}_2(0) =
\tilde{N}_8(0)=1/8$ and $\tilde{N}_1(0) = \tilde{N}_4(0) =
\tilde{N}_7(0)=1/4$; Panel~(c): $\tilde{N}_3(0) = \tilde{N}_5(0) =
\tilde{N}_6(0)=0$, $\tilde{N}_1(0) = \tilde{N}_7(0) = 1/8$ and
$\tilde{N}_2(0) = \tilde{N}_4(0) = \tilde{N}_8(0)=1/4$ where
$\tilde{N}_i(t) = N_i(t)/N_{tot}$, $i=1,\ldots,8$. The initial
conditions are also illustrated in the insets: a darker shading
indicates that more particles are injected into the container.}
\label{NetworkPanel}
\end{figure}

\end{document}